\documentclass[letterpaper,11pt]{article}
\usepackage{jheppub}
\usepackage{braket}
\usepackage{tikz}
\usepackage{hyperref}
\usepackage{amssymb}
\usepackage{amsmath}
\usepackage{color}
\usepackage{amsthm,amsfonts,graphicx,verbatim,color,cancel}
\DeclareMathOperator{\tr}{tr}

\newcommand{\nn}{\\ \nonumber}

\newcommand \mathtikz[1] {\quad \vcenter{\hbox{\tikz{#1}}} \quad}
\newcommand{\pd}{\partial}

\newcommand\pairA[2]{ 
\begin{scope}[xshift=#1,yshift=#2]
\draw (-0.75,0) -- (-0.25,0) to [out=-90,in=180] (0,-0.33) to [in=-90,out=0] (0.25,0) -- (0.75,0) to [out=-90,in=0] (0,-0.83) to [out=180,in=-90] (-0.75,0);
\end{scope}
}

\newcommand\copairA[2]{ 
\begin{scope}[xshift=#1,yshift=#2]
\draw (-0.75,0) -- (-0.25,0) to [out=90,in=180] (0,0.33) to [in=90,out=0] (0.25,0) -- (0.75,0) to [out=90,in=0] (0,0.83) to [out=180,in=90] (-0.75,0);
\end{scope}
}

\newcommand\deltaA[2]{ 
\begin{scope}[xshift=#1,yshift=#2]
\draw (-0.75,-1) -- (-0.25,-1) to [out=90,in=180] (0,-0.66) to [in=90,out=0] (0.25,-1) -- (0.75,-1) to [in=-90,out=90] (0.25,0) -- (-0.25,0) to [in=90,out=-90] (-0.75,-1);
\end{scope}
}

\newcommand\zipper[2]{ 
\begin{scope}[xshift=#1,yshift=#2]
\draw (-0.25,-1) -- (0.25,-1);
\filldraw[right color=white,left color=lightgray] (-0.25,0) to (-0.25,-1) to [out=90,in=225] (0,-0.5) to [out=-45,in=90] (0.25,-1) to (0.25,0);
\filldraw[left color=white,right color=lightgray] (0,0) ellipse (0.25 and 0.1);
\end{scope}
}

\begin{document}

\title{Entanglement branes and factorization in conformal field theory}
\affiliation[a,b]{Department of Physics, Fudan University, 2005 Songhu Road, 200438 Shanghai, China}

\affiliation[a]{State Key Laboratory of Surface Physics
Fudan University,
200433 Shanghai, China
}
\affiliation[a]{Collaborative Innovation Center of Advanced Microstructures,
210093 Nanjing, China
}
\affiliation[a]{Department of Physics and Center for Field Theory and Particle Physics,
Fudan University
200433 Shanghai, China
}
\affiliation[a]{
Institute for Nanoelectronic Devices and Quantum computing,
Fudan University,
200433 Shanghai , China}
\author[a]{Ling Yan Hung}
\emailAdd{lyhung@fudan.edu.cn}
\author[b]{Gabriel Wong}
\emailAdd{gabrielwon@gmail.com}

\abstract{In this work, we consider the question of local Hilbert space factorization in 2D conformal field theory.   Generalizing previous work on entanglement and open-closed TQFT, we interpret the factorization of CFT states in terms of path integral processes that split and join the Hilbert spaces of  circles and intervals.  More abstractly, these processes are cobordisms of an extended CFT  which are defined purely in terms of the OPE data. In addition to the usual sewing axioms,  we impose an entanglement boundary condition that is satisfied by the vacuum Ishibashi state.   This choice of entanglement boundary state leads to reduced density matrices that sum over super-selection sectors, which we identify as the CFT edge modes.  Finally, we relate our factorization map to the co-product formula for the CFT symmetry algebra, which we show is equivalent to a Boguliubov transformation in the case of a free boson.

}

\maketitle

\section{Introduction}
In its usual path integral formulation, a continuum QFT does not come equipped with a notion of local Hilbert space factorization.   This presents an obstruction to defining the reduced density matrix for a subregion which presumes such a factorization and a well defined partial trace.   In the algebraic formulation of QFT, path integrals,  Hilbert space factorization, and reduced density matrices are  eschewed all together.  Instead one assigns algebras of local operators to sub-regions of spacetime and quantum information theoretic quantities such as relative entropy are defined as properties of these local algebras.    While the Algebraic approach has it's conceptual appeal and holds the promise of rigor,  the calculational tools and the number of models that can be formulated in this way remains limited. On the other hand, the extended Hilbert space approach offers a useful alternative that falls within the standard paradigm of QFT.  In previous work, we applied the framework of 2D extended TQFT to give local constraints on the Hilbert space extension \cite{Donnelly:2018ppr}.  The goal of this paper is to generalize this construction to two dimensional CFT's.
 
The extended Hilbert space construction is best understood in gauge theory \cite{Donnelly:2014gva} \cite{Buividovich2008b}, where the correlations introduced by the Gauss law constraint prevents factorization of the gauge invariant Hilbert space into independent local factors.   Instead, each local Hilbert space is extended to include boundary edge modes,  which transform nontrivially under the gauge group, now interpreted as a boundary symmetry group.  The physical Hilbert space is then recovered as the \emph{entangling product} of the independent factors, a fusion product which projects onto the gauge invariant subspace where the Gauss law is satisfied \cite{Donnelly:2016auv}.   This extension leads to an edge mode contribution to the entanglement entropy of gauge theories; in the case of 2+1 D topological gauge theories this explains the topological entanglement entropy which plays a central role as an order parameter for topological order. The factorization problem is even more acute in quantum gravity, where it  has been suggested that edge modes associated with diffeomorphisms may underly the puzzle of Bekenstein Hawking entropy \cite{Donnelly:2016auv}.  In the case of 2D JT gravity, the factorization problem was analysed in \cite{Jafferis:2019wkd}

\begin{figure}[h]
\centering
\includegraphics[scale=.5]{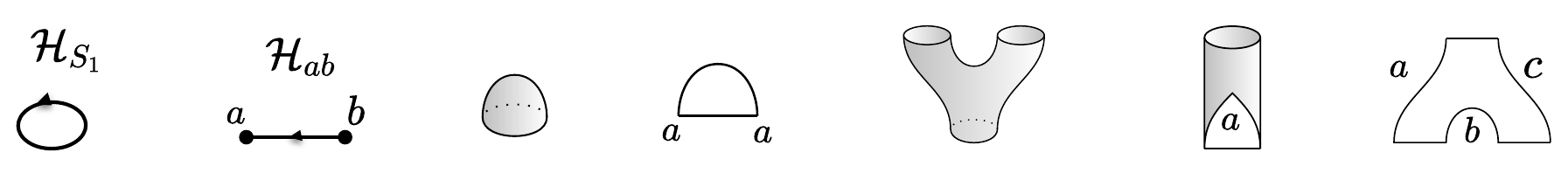}
\caption{The basic objects  and cobordisms of a 2D extended TQFT. The extension from the closed to open sector introduces boundary labelings on the open string, which has to match whenever their endpoints meet}
\label{BC}
\end{figure}

\begin{figure}
\centering
\includegraphics[scale=.5]{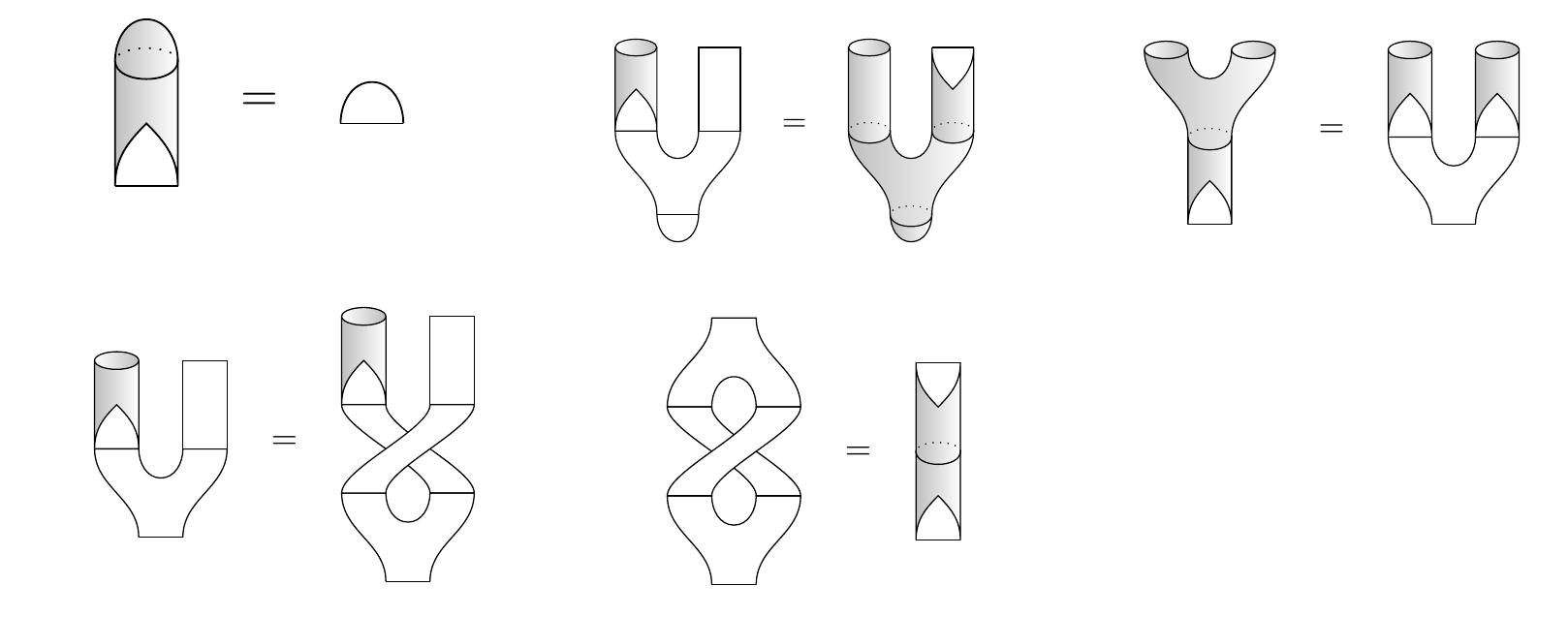}
\caption{Shown are five sewing axioms which ensures the consistency of different gluings. The last one is the Cardy condition}
\label{Msegal}
\end{figure}

  For 2D gauge theories and TQFT's, we gave an axiomatic formulation of the Hilbert space extension in the framework of \emph{extended} TQFT \cite{Donnelly:2018ppr}.   This is a categorical description of TQFT in which the path integral is viewed as a rule which assigns Hilbert spaces to codimension 1 manifolds, and linear maps to cobordisms which interpolates between these manifolds.
Since gluing cobordism corresponds to composition of linear maps, the path integral on an arbitrary surface can be constructed from gluing a basic set of cobordisms (figure \ref{BC}).   These are subject to sewing relations that ensure the consistency of different gluings (figure \ref{Msegal}) \cite{Moore:2006dw}. 

In 2D, the cobordisms are worldsheets, and the extended TQFT is organized into a closed string sector and open string sector corresponding to closed and open manifolds. Starting with the closed sector, we extend the TQFT to the open sector by introducing boundary labels for the open string, and then define open Hilbert spaces and open-closed cobordisms.    In \cite{Donnelly:2018ppr}, we incorporated entanglement into this framework by interpreting the factorization maps (figure \eqref{ebrane}) as elements of the basic open-closed cobordisms of the extended TQFT . 
Thus the extension of the Hilbert space is identified with the extension of the closed TQFT, and the boundary labels coincide with the edge modes in the extended Hilbert space.  These edge modes must satisfy the \emph{locality} constraints expressed by the sewing axioms.  In addition, for the factorization to preserve the correlations in the original state, this extension must satisfy the E brane axiom \cite{Donnelly:2018ppr} which requires that all holes created by the evolution of an entangling surface to be closed up.  This is equivalent to the statement that the reduced density matrix $\rho_{V}$ of a subregion $V$ should reproduce all correlation functions operators $O_{V}$ :
\begin{align}\label{O}
\braket{O_{V}} = \tr_{V} (\rho_{V} O_{V}) 
\end{align}
In addition to providing a solution to the factorization problem, the cobordism approach also gave a systematic way to compute mult-interval modular flows, negativity, and entanglement entropy \cite{Donnelly:2018ppr}.

In this work we attempt a generalization of the extended TQFT approach to Hilbert space factorization to 2D conformal field theories.   In this case, the basic cobordisms are specified by the OPE coefficients  that define a boundary CFT (figure \eqref{OPE}).  Just as in the TQFT case, these are constrained by well known sewing axioms\cite{Lewellen:1991tb} of the BCFT, the most well-known of which is crossing symmetry and modular invariance.   Following the analogy with extended TQFT,  we will formulate an E brane axiom for CFT edge modes \eqref{EB} and use it to define a Hilbert space factorization consistent with OPE's.  We will show that a factorization map consistent with \eqref{O} requires a sum over superselection sectors labelled by Cardy boundary conditions.  Since each Cardy boundary condition satisfies the BCFT sewing axioms, the sum over these superselection labels automatically satisfies them as well.   

Here is a brief summary of the paper:  In section 2 we review the basic cobordism of an extended CFT \cite{Lewellen:1991tb}.  In section 3 we formulate an E brane constraint  \eqref{EB} for CFT's and propose a solution in the form of the vacuum Ishibashi state. We explain how this choice is dictated by conformal symmetry, and corresponds to summing over local boundary conditions \eqref{eb} in contrast to the factorization maps previously proposed\footnote{The authors in \cite{Ohmori:2014eia} were considering different problem of how to define a CFT factorization that reproduces a particular choice of factorization on a spin chain.  In this work, we ask the question of what is the right factorization for the CFT that preserves equation \eqref{O} for all local operators. These two questions have different answers, but there is no contradiction; rather there is a small mismatch in the definition of locality in the spin chain compared to the CFT, so some local operators in the CFT will not be local in the spin chain. } \cite{Ohmori:2014eia}.   As an illustration, we consider the vacuum state of the free compact boson to give an explicit example of the E brane, the associated CFT edge modes, and the edge entanglement entropy.  In this case the sum over boundary conditions restores the $U(1)$ symmetry of the theory and cancels an anomalous term in the entropy that would appear for a fixed boundary condition.  We then propose a factorization map \eqref{V} in terms of three point functions of a general CFT and apply it to the primaries of the compact boson.   In section 4 we interpret the factorization map in terms of the fusion rules and the ring-like tensor products of Verma modules \cite{Gaberdiel:1993td}.   This allows us to formulate the factorization of descendant states using a co-product formula which is dictated by the relevant symmetry algebra\footnote{In this paper we make use of the co product formula associated with the Kacs-Moody algebra, but a similar formula applies to the Virasoso algebra and can be used to define factorization in a general CFT}. Together with the results of section 3, this gives a factorization map for a complete basis of states purely in terms of CFT data.  To gain some intuition for this co-product factorization \eqref{Delta} we make a comparison with the standard Bogoliubov transformation \eqref{bog} which underlies the Unruh effect in Minkowski space, and find a highly non-trivial match.  We will conclude with some discussion of open problems and possible applications of our ideas to  tensor network renormalization  and ``BC bits" formulation of holographic CFT's \cite{VanRaamsdonk:2018zws}.

\section{ Basic data of an open-closed CFT}
Here we present the boundary CFT data in the spirit of extended QFT. 

\paragraph{Local Hilbert spaces } In the closed CFT, the codimension 1 objects are circles, which are assigned to a Hilbert space furnishing a representation of the Virasoro algebra.  In addition we will assume the presence of a Kac-Moody symmetry algebra and that states are classified according to it's representations, labelled by a primary state $\ket{\phi_{i}}$ (For brevity we will consider a chiral sector of the closed CFT. )  
The entire representation $\mathcal{H}_{i}$  is obtained from applying the negative modes of the Kac-Moody algebra to $\ket{\phi_{i}}$.  For states defined around the origin, these are obtained from the Kac-Moody current $J(w)$ by 
\begin{align}
J^{a}_{n} &= \oint_{w=0} w^n J(w) dw \nn
\mathcal{H}_{i} &= \text{span}\{ J^{a}_{-n_{1}} \cdots J^{a}_{-n_{k}} \ket{\phi_{i}} \}
\end{align}  
Note that in contrast to a generic QFT where Hilbert spaces are assigned to global Cauchy slices, in a CFT there exists naturally a notion of a \emph{local} Hilbert space.
This is due to the state operator correspondence, which says that we can define a Hilbert space $\mathcal{H}_{i}(z) $ around an arbitrary puncture $z$ on a Riemann surface by inserting a primary operator $\phi_{i}(z) $.   Around each puncture, we can build a tower of states via the modes of the current centered at the puncture: 
\begin{align}\label{tower} 
J^{a}_{n}(z)  &= \oint_{z} (w-z)^n J^{a}(w) dw \nn
\mathcal{H}_{i}(z) &= \text{span}\{ J_{-n_{1}}(z) \cdots J_{-n_{k}}(z) \ket{\phi_{i}} \}
\end{align}  

The closed sector is extended to the open sector by introducing intervals $I_{ab}$ with boundary labels $a,b$ specifying conformally invariant boundary conditions.   The corresponding Hilbert space $\mathcal{H}_{ab}$ is again locally defined around a puncture at the \emph{boundary} of an open surface.  At each puncture we insert a boundary condition changing operator $\psi_{i}^{ab}$ corresponding to a highest weight state, upon which the representation $\mathcal{H}^{i}_{ab}$ is built by applying the Kac-Moody raising operators.  We will assume the open string can be ``unfolded" into a chiral half of a closed string, and the Hilbert space $\mathcal{H}^{i}_{ab}$ corresponds to a representation of a single chiral copy of the Kac-Moody algebra.  As discussed below, this will be the case for conformally invariant boundary conditions that preserve exactly one chiral copy of all other global symmetries as well.

\begin{figure}[h]
\centering
\includegraphics[scale=.37]{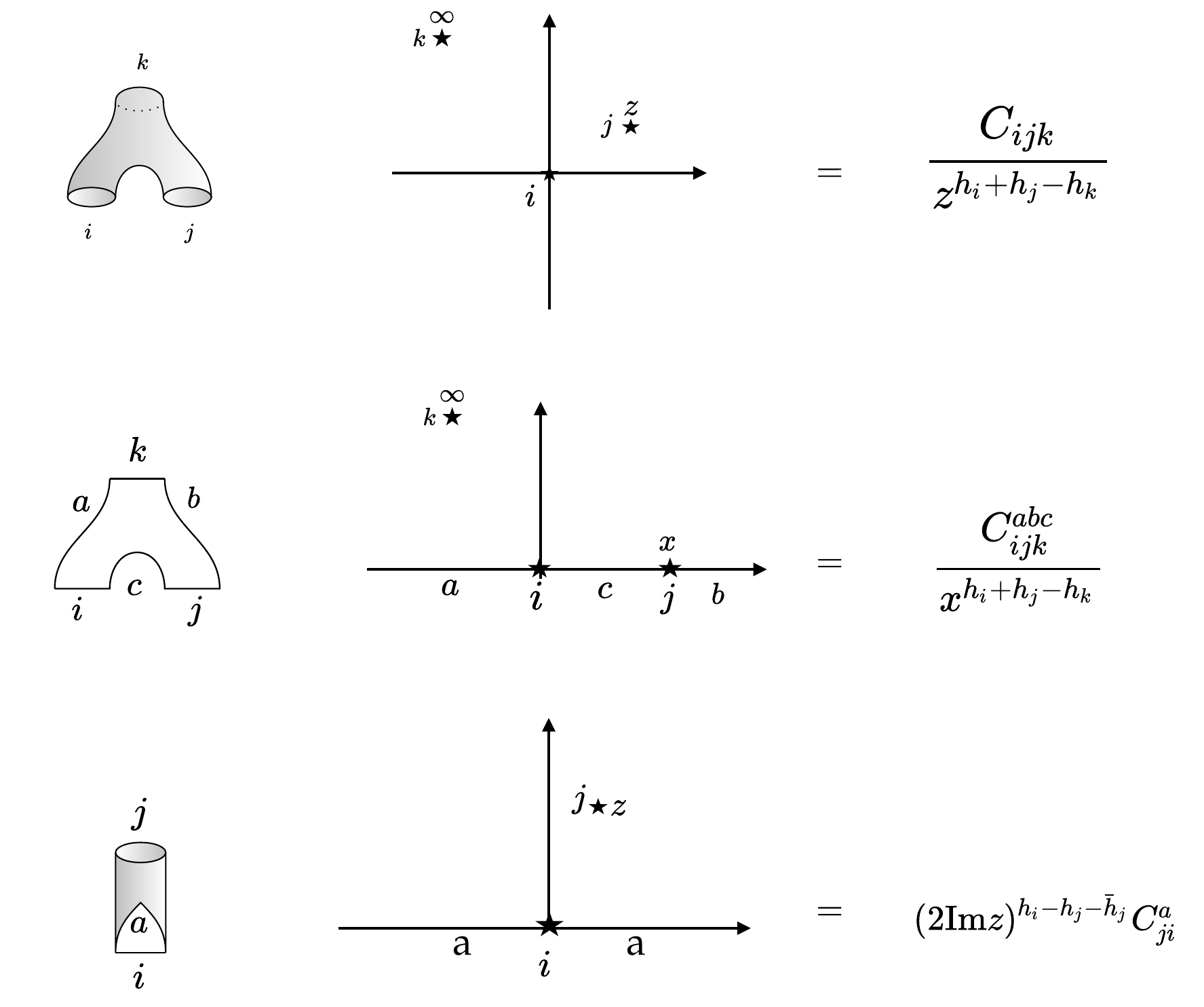}
\caption{ The basic CFT cobordisms can be mapped to the punctured (half) plane. Given a choice of punctures, the cobordism is a linear map between Hilbert spaces at the punctures, whose components are determined by the OPE coefficients.}
\label{OPE}
\end{figure}

\paragraph{The basic interaction cobordisms}
The basic cobordisms of an extended CFT are defined by the OPE coefficients, and are summarized in figure \ref{OPE}.  On the upper half z-plane, the three types of OPE's are
\begin{align}
\phi_{i}(z) \phi_{j}(0) &\sim \sum_{i,j,k}  \frac{C_{ijk}}{z^{h_{i}+h_{j}-h_{k}} \bar{z}^{\bar{h}_{i} +\bar{h}_{j} -\bar{h}_{k}}} \phi_{k}(z)+ \cdots\nn
\psi^{ab}_{i}(0)\psi^{bc}_{j}(x) &\sim \sum_{k}  C^{abc}_{ijk} x^{h_{k}-h_{j}-h_{k}} \psi^{ac}_{k}(x)  + \cdots \nn
\phi_{j,\bar{j}}(z) &\sim \sum_{i}(2 \text{Im} z)^{ h_{i}-h_{j} -\bar{h}_{j}} C_{ji}^{a} \psi_{i}^{aa}(\text{Re }z) + \cdots  ,
\end{align}
where $\phi_{i}$, $\psi^{ab}_{i}$, are bulk and boundary primaries respectively. The last equation is the bulk-boundary OPE describing the effect of bringing a bulk field close to the boundary \footnote{The leading term in this OPE expansion $C_{i\bar{i} ,0}^{a}$ is given determined by the one point functions:
\begin{align}
\braket{1}_{a} &= A^{a}_{00} \nn
\braket{ \phi_{i, \bar{i}} (z,\bar{z}) } &= \frac{A_{i,\bar{i}}^{a} }{(z-\bar{z})^{h_{i}-\bar{h}_{i} }}\nn
C_{i\bar{i} ,0}^{a} &= \frac{ A_{i\bar{i} } ^{a}}{A_{00}^{a}}
\end{align} 
}

In this work we introduce factorization maps for the interval and the circle given by the cobordism for open string splitting and the closed to open transition.    These will differ from the standard definitions in figure \eqref{OPE} by the choice of entanglement boundary conditions, which we elaborate on below.  Moreover we will make a choice of conformal frame where the factorized states live on large intervals rather than at punctures, as shown in the left of figure \eqref{cp4}. This is the choice compatible with standard entanglement calculations.      
\paragraph{Conformally invariant boundary conditions and Sewing Relations} 
Here we give a brief review of conformally invariant boundary conditions and the relevant sewing relations.  On the radially quantized upper half plane, these are boundary conditions on the real line which satisfy 
\begin{align}\label{TT}
    T(z) = \bar{T}(\bar{z}) \quad z \in \mathbf{R} 
\end{align}
Physically, this condition requires that no momentum flows out of the boundary.  It cuts down the full Virasoro symmetry  in to a single chiral copy, corresponding to conformal transformation preserving the boundary. Due to the Sugawara condition $ T(z) \sim: J(z) J(z): $ we can satisfy this relation by solving \footnote{More generally, along the real line,  $\bar{J}(\bar{z})$ is related to $ J(z) $  by an automorphism of the symmetry algebra}
\begin{align}\label{CB}
 J(z) =  \pm \bar{J}(\bar{z}) \quad z \in \mathbf{R}.
\end{align}
In other words, we are restricting to situations in which one copy of the Kac-Moody algebra of the RCFT is
preserved at the boundary. This collection of boundary conditions, as we are going to see, suffices in satisfying the E-brane axiom. 
For example, \eqref{CB} can be solved in diagonal (Rational) CFT's and for the compact boson.  These relations allow the charges and the states created by them to be unfolded into the lower half plane by defining
\begin{align} 
J(z)& = \pm \bar{J}(\bar{z}^*)  \quad \Im z <0\nn
J_{n} &= \int_{-\pi}^{\pi} dz\, z^{n} J(z) \mp \int_{-\pi}^{\pi}  d \bar{z} \,\bar{z}^n \bar{J}(\bar{z})= \oint dz\, z^{n} J(z)
\end{align} 
In the second equation, we combined the holomorphic and anti-holomorphic charges on the upper half plane into a single holomorphic charge in the whole plane.  The open string states thus transform in a single chiral sector of the symmetry algebra. 

\paragraph{ Ishibashi states, physical boundary states and cardy condition} 
\begin{figure}[h]
\centering
\includegraphics[scale=.45]{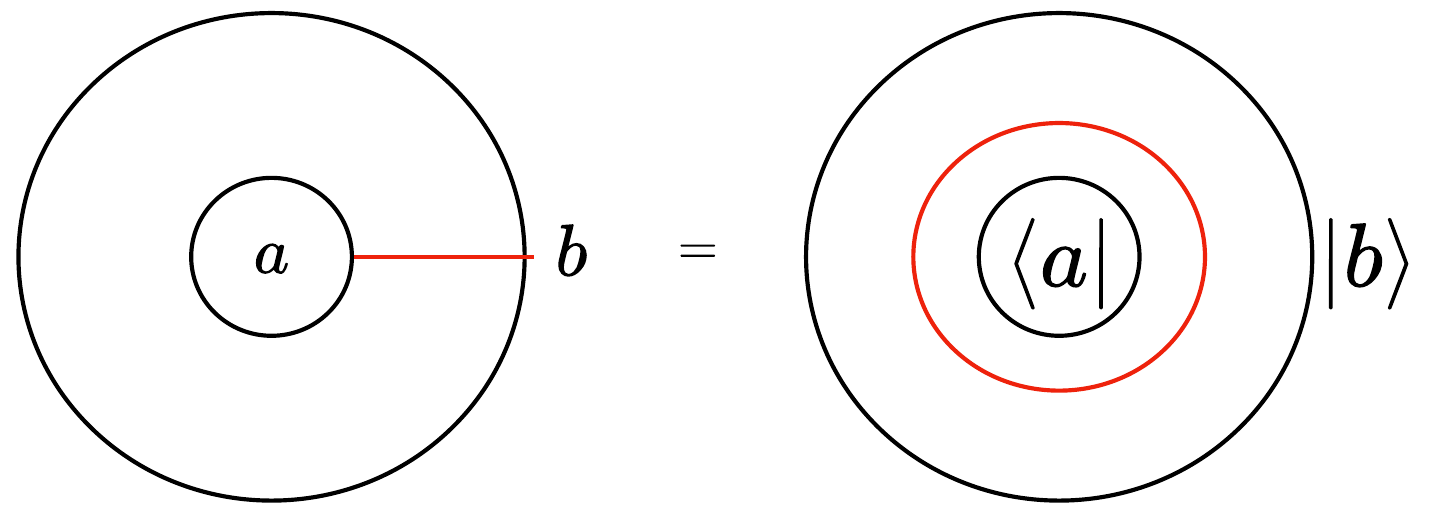}
\caption{By viewing the annulus path integral as a closed string amplitude, the open string boundary conditions can be mapped to closed string states. }
\label{cardy}
\end{figure}

 As shown in figure \eqref{cardy}, we can describe open string boundary conditions $a,b$ on an interval $I_{ab}$ by evolving it in a close loop, and then treating the resulting cylinder as an amplitude between closed string boundary states.  In the closed string picture, the conformal invariance condition \eqref{CB} can be expressed as 
\begin{align} \label{B}
   ( J_{n}\pm \bar{J} _{-n})\ket{B}=0
\end{align}
where the Kac-Moody charges now live on the circles trace out by the open string endpoints.  The Ishibashi states form a complete set of boundary states that satisfy \eqref{B}.  These are left-right entangled closed string states labelled by bulk conformal dimensions $h,\bar{h} $ : 
\begin{align}
    |h\rangle\rangle =  \sum_{N} |h,N\rangle \otimes|\bar{h},\bar{N}\rangle
\end{align}
Here $N$ is a label for the descendants.
Even though the Ishibashi states are conformally invariant due to \eqref{CB}, they do not necessarily correspond to physical boundary conditions for the open string; in particular they do not have to be \emph{local}.  The physical boundary conditions must satisfy further non-linear constraints determined by the sewing relations alluded to in the introduction. These were first formulated for boundary CFT in \cite{Lauda:2005wn} and they ensure that different gluings of the elementary cobordisms (figure \eqref{OPE}) into the same manifold will give the same linear map.  For a given closed CFT, we can view an extension to the open sector as a solution to these constraints. 

The constraint we will focus on is the Cardy condition, illustrated in figure \eqref{cardy}.  This requires the physical boundary state $\ket{a},\ket{b}$ inserted on the annulus gives rise to bonafide open string partition functions that sums over an open string Hilbert space $\mathcal{H}_{ab}$: 
\begin{align}
   \langle a| e^{-T H_{\text{closed}}} |b \rangle = \text{tr}_{ab} e^{-\frac{1}{T} H_{open}}
\end{align}
This places strong constraints on $\ket{a,b}$ because the RHS must be a sum over Boltzmann factors with \emph{integer} degeneracies.  The solutions are called Cardy states, which can be expressed as particular linear combinations of the Ishibashi states.  However note that arbitrarily superposing Cardy states does not return a Cardy state, since the Cardy constraint is non linear.

\section{The E brane boundary condition and the vacuum Ishibashi state }
\begin{figure}[h]
\centering
\includegraphics[scale=.45]{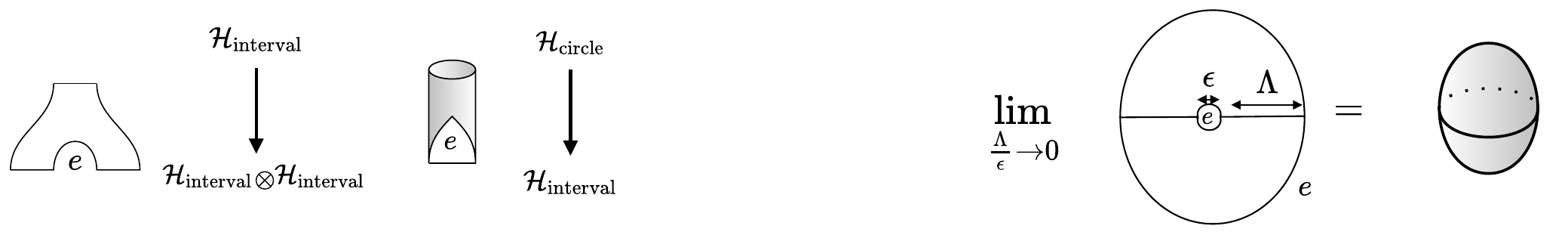}
\caption{We define factorization maps from the open-closed cobordisms on the left. These satisfy the an E-brane boundary condition that allows us to close up holes.  }
\label{ebrane}
\end{figure}

In this work, we wish to extend a closed CFT by cobordisms in figure \eqref{ebrane} that provide a factorization of states on the interval or circle.  Such a factorization map and the associated entanglement boundary conditions should preserve the correlation function of the original state. 
Following the analogous construction in a TQFT \cite{Donnelly:2018ppr},  we formulate this boundary condition via a closed string boundary state which satisfies a constraint we call the entanglement brane, in addition to the sewing axioms.  For a CFT, we propose an E brane axiom which requires that as the entangling surface shrinks to zero size, the bulk correlation functions should approach their value in the absence of the boundary.   For example, for an annulus with two entanglement boundaries of size $\epsilon$ and $\Lambda$ , we require that 
\begin{align} \label{EB}
\lim_{q=\frac{\Lambda}{\epsilon} \rightarrow 0} \langle \phi (z_{1})\cdots \phi(z_{n})\rangle_{q}  = \langle \phi (z_{1})\cdots \phi(z_{n})\rangle_{S^2 }
\end{align}
Conformal invariance then implies that the \emph{vacuum} Ishibashi is the E brane boundary state. 

To understand what this boundary state implies in the open channel, consider theories with a complete set of physical boundary states $\mathcal{B}$ satisfying the sewing axioms,and therefore correspond to \emph{local} boundary conditions.  Completeness means we can express  $|a\rangle\in \mathcal{B} $ in terms of the Ishibashi states  \cite{Pradisi:1996yd} via an invertible matrix:
\begin{align}
\ket{a}  &= \sum_{i} \frac{K_{a i}}{(S_{0i}^{1/2})}  | i \rangle \rangle \nn
\sum_{a \in \mathcal{B} } K_{ai} K_{aj} &= \delta^{ij},
\end{align} 
The second equation allows us to invert the first equation to write Ishibashi states as linear combinations of the \emph{physical}  boundary states.  In particular,the E brane boundary state is 
\begin{align}\label{eb}
|e\rangle \rangle = |0\rangle\rangle = \sum_{a}  (S_{0i}^{1/2}) K_{a0} \ket{a}
\end{align} 
For a diagonal RCFT, the physical boundary states are the Cardy states, and $K_{ai}=S_{ai}$ are the S-matrix elements.  Thus we see that the E brane boundary state correspond to a \emph{sum over local boundary conditions}. Since the Cardy conditions are non linear, the Cardy states do not form a vector space.   Instead, we should interpret \eqref{eb} as a sum over super-selection sectors labelled by Cardy boundary conditions.  The fact that the cardy state in each superselection sector satisfies the sewing axioms then implies the E brane state does as well.  

The reason the E brane must involve a sum over cardy state as in \eqref{eb} can be understood by applying the E brane axiom to the identity operator on plane with a small hole.  The E brane axiom implies this should equal the path integral on a sphere, which we normalize as 
\begin{align}
\braket{1}_{S^{2} }=1 
\end{align} 
Expanding our E brane boundary state in terms of Cardy state, 
\begin{align}\label{e}
    |e\rangle\rangle  &= \sum_{a}  c_{a}\ket{a}
\end{align}
the E brane axiom becomes 
\begin{align}
\braket{1}_{S^{2} }&= \langle0|e\rangle\rangle= \sum_{a} c_{a} \braket{0|a} =1
\end{align}
We can solve for the constants $c_{a}$ in an RCFT by noting that 
\begin{align}
      \braket{0|a} =  \frac{S_{a 0}}{\sqrt{S_{00}}}
\end{align}
A solution is given by $c_{a} =(S_{0i}^{1/2}) S_{a0} $ leading to the Vacuum Ishibashi state in \eqref{eb}.  This calculation works for an arbitrary number of small entanglement boundaries, since a neighborhood of each boundary can be conformally mapped to a long cylinder which projects to the vacuum part of the boundary state.
From this argument, we see that the Vacuum Ishibashi state satisfies the E brane axiom because it is the conformally invariant boundary state which has trivial overlap with the vacuum.   The same argument applies when there are bulk operator insertions away from the entangling surface, since we have schematically 
\begin{align}
    \braket{e|\phi_{1}\cdots \phi_{n}  |e} \sim  \braket{e|0}\braket{0|e^{-\frac{H}{\epsilon} } \phi'_{1}\cdots \phi'_{n} e^{-\frac{H}{\epsilon}} |0  } \braket{0|e} =  \braket{0| \phi'_{1} \cdots \phi'_{n} |0  },
 \end{align}
where  $\phi'$ are the conformally transformed fields.  

Another indication that a sum like \eqref{eb} is needed for the E brane boundary state comes from the following observation. For the case of a compact boson, a fixed Cardy state corresponding to a Dirichlet or Neumann  boundary leads to an anomalous term  $\log \log \frac{L}{\epsilon} $ in the entanglement entropy of a single interval of length $L$ \cite{Michel:2016fex}.    As we show below, the vacuum Ishibashi state leads to a sum over edge modes ( in this case Dirichlet or Neumann boundary conditions) that cancels this anomalous term.   
\paragraph{Sum over selection sectors, edge and bulk entanglement entropy}
The E brane extension defines a reduced density matrix that involves a direct sums over super selection sectors labelled by physical boundary conditions.  In the context of 2+1 D scalar fields, these superselection sectors have been discussed in \cite{Agon:2013iva}, and their corresponding edge modes analysed in \cite{Campiglia:2018see}.  In the algebraic formulation, the presence of superselection sectors is due to appearance of a nontrivial center in the algebra assigned to a subregion \cite{Lin:2018bud, Casini:2013rba} , and are labelled by the eigenvalues of the center variables.  
 
Given a generic label $k$ for the super selection sectors with the unnormalized  density matrix $\tilde{\rho}_{k}$, we can write the total unnormalized density matrix and it's trace as
\begin{align}
\tilde{\rho} &=  \bigoplus_{k}   \tilde{\rho}_{k} \nn 
Z&\equiv \tr \tilde{\rho} =\sum_{k} \tr \tilde{\rho_{k}}
\end{align} 
We define the normalized density matrix $\rho_{k}$ in the $k$ th sector and normalized probabilities $P_{k}$ by
\begin{align}
\rho_{k}&\equiv\frac{\tilde{\rho}_{k}}{Z_k} , \qquad Z_k \equiv \tr(\tilde \rho_k)\nn
P_{k}&\equiv \frac{\tr \tilde{\rho}_{k}}{Z}
\end{align} 
so that the total normalized density matrix is
\begin{align}
\rho = \bigoplus_{k}  P_{k} \rho_{k}
\end{align}
Since the trace commutes with the direct sum, the total entanglement entropy naturally splits into a ``edge mode" Shannon entropy plus a bulk piece 
\begin{align}  \label{eq:EE}
S&=- \sum_{k} P_{k} \log P_{k} + \sum_{k} P_{k} S_{k} \nn
S_{k} &= - \tr \rho_{k} \log \rho_{k} 
\end{align}
In particular for a single interval with E brane extension, we have super selection labels $k=a \,b$, with $a, b\cdots$ specifying local boundary conditions at the two ends. 

We can then define a reduced density matrix in each sector by 
\begin{align}
\tilde{\rho}_{ab} =  \bigoplus_{ab} \langle\langle0\ket{a}  \bra{b}0\rangle \rangle  e^{-H_{ab}/T} 
\end{align} 
with $H_{ab}$ the open string Hamiltonian in each sector, and $T$ is the length of the interval\footnote{In an abuse of notation we have denoted the expansion coefficient in \eqref{e} by  $\bra{a}0\rangle \rangle=\frac{S_{a0}}{\sqrt{S_{00}}}$ even though we are not using the Hilbert space inner product because the boundary states are not normalizable. }. This is defined so that the trace is 
a path integral with Ishibashi states inserted: 
\begin{align}
Z= \sum_{ab} \tr \tilde{\rho}_{ab}  &=\sum_{ab} \langle\langle0\ket{a}  \bra{b}0\rangle \rangle  \tr e^{-H_{ab}/T}  \nn
&=  \langle\langle 0|e^{-TH_{\text{closed}}} |0 \rangle\rangle 
\end{align} 
Note that when computing the entropies using the replica trick, we do not insert the Ishibashi onto the replicated geometry.   Instead, the replicated density matrix is
\begin{align}
\tr \tilde{\rho}^{n} = \sum_{ab} \langle\langle0\ket{a}^{n}  \bra{b}0\rangle \rangle^n  \tr e^{-n H_{ab}/T} 
\end{align}
which is different than $ \langle\langle 0|e^{-\frac{T}{n} H_{\text{closed}}} |0 \rangle\rangle $ 

Defining $Z_{ab}= \tr e^{-H_{ab}/T}= \langle a |  e^{-T H_{\text{closed}}} | b \rangle $, we can use the closed string expressions 

\begin{align}
\tr \tilde{\rho}_{ab}&= \ \langle\langle0| a \rangle Z_{ab} \langle b | 0\rangle\rangle \nn
P_{ab} &=\frac{ \langle\langle0| a \rangle Z_{ab} \langle b | 0\rangle\rangle } {Z}
\end{align}
to compute the entropies. In particular the edge entropy takes the thermal form 
\begin{align}
S_{\text{edge}} &=\braket{ F}- \log Z\nn
F(ab)&=  \log  \ \langle\langle0| a \rangle  + \log  \langle b | 0\rangle\rangle +\log       Z_{ab} 
\end{align}
where the average $\braket{\cdots}$ is with respect to $P_{ab}$.  \subsection{E brane for the free compact Boson} 
In this section, we illustrate the ideas above by explicitly constructing the E brane extension for the free compact boson.  
In Minkowski signature, various choices of entanglement boundary conditions for this system was analysed in \cite{Michel:2016fex} , who also calculated the entanglement entropy of the associated edge modes. Below we offer a different point of view based on Euclidean cobordisms and the E brane boundary state. 

Consider a free boson\footnote{Our normalization corresponds to $\alpha' = \frac{1}{2}$} of radius $R$:
\begin{align}
S&= \frac{1}{2\pi } \int dx dt (\pd_{t}\phi)^2  +  (\pd_{x}\phi)^2\nn
\phi &= \phi + 2 \pi R 
\end{align} 
In the closed string sector, the zero modes are parametrized by left/rightmomenta $p_{R},p_{L}$
\begin{align}
    \ket{p_{L},p_{R}}= \ket{ \frac{m}{2R} +n R, \frac{m}{2R}-nR}  \quad m,n \in \mathbb{Z}
\end{align}
where $m$ counts the units of momentum and $n$ the winding number.   
In terms of the right/left oscillators  $\alpha_{n},\tilde{\alpha}_{n}$ that generate the $U(1)$ Kac-Moody algebra, there are two sets of Ishibashi states
\begin{align}
|nR,-nR \rangle \rangle     &= \exp{\sum_{l=1} \frac{- \alpha_{-l} \tilde{\alpha}_{-l}}{l}} \ket{nR,-nR}\nn
|\frac{m}{2R},\frac{m}{2R} \rangle \rangle     &= \exp{\sum_{l=1} \frac{- \alpha_{-l} \tilde{\alpha}_{-l}}{l}} \ket{\frac{m}{2R},\frac{m}{2R}}
\end{align}
Although the free boson is not a rational CFT, the Cardy states are known and given as follows:
\begin{align}\label{DN}
  \textrm{Neumann:}   \,\, ||\tilde{\phi}_{0}  \rangle\rangle &= R^{1/2} \sum_{n\in \mathbb{Z}} e^{i \tilde{\phi}_{0} n R}|nR,-nR \rangle \rangle  \nn
 \textrm{Dirichlet:}  \,\, ||\phi_{0}\rangle\rangle &= (2R)^{-1/2}\sum_{m \in \mathbb{Z}} e^{i \frac{ m \phi_{0}}{R}}  |\frac{m}{2R},\frac{m}{2R} \rangle \rangle. 
\end{align} 
In the open string frame,   $\phi_{0}$ is the Dirichlet boundary value of $\phi$  and $\tilde{\phi}_{0} $ is a ``Wilson line" variable that is best understood as the T-dual variable to $\phi_{0}$.  The E brane boundary state is the vacuum Ishibashi  state $|0,0\rangle\rangle$, which can be explicity obtained by summing over Dirichlet or Neumann cardy states: 
\begin{align} \label{D}
\ket{e}=|0,0 \rangle \rangle &=  \pi \sqrt{2 R}\int_{0}^{2\pi R}  d\phi_{0}   || \phi_{0}\rangle \rangle\\
&=\frac{2 \pi }{\sqrt{R} }  \int_{0}^{\frac{\pi}{R}}  d \tilde{\phi}_{0}  ||  \tilde{\phi}_{0} \rangle \rangle \label{N}
\end{align}

The sum over Dirichlet boundary condition is natural  because it is needed to preserve the shift invariance of $\phi$, which enforce momentum conservation\footnote{This is preserved exactly in the vacuum Ishibashi state, where as a cardy state will preserve the shift invariance up to order $\epsilon$ } in the correlation functions of vertex operators $e^{i n \phi} $ .  It also makes sense from the point of view of the path integral, which sums over all fluctuations of $\phi$ at the entangling surface \cite{Headrick:2012fk}.   It can be checked explicitly that the sum over Dirichlet boundary condition satisifies \eqref{EB} for all \emph{local} correlation functions.  For the Neumann E brane, the reason for summing over Wilson lines can be understood via T-duality.  Even though shift invariance for correlators of \emph{ local } operators are preserved for any fixed Neumann boundary condition, we have to ensure that the same holds for correlators of \emph{non-local} vortex creation operators.  Under T-duality mapping $\phi \rightarrow  \tilde{\phi}$ with dual radius $\tilde{R} = \frac{2}{R} $,  these non local operators are mapped to local operators of the form $V_{m} = \exp(i \tilde{n} \  \tilde{\phi }) $. Shift invariance for correlators of these operators require a sum over the T-dual dirichlet boundary condition, $\tilde{\phi}_{0} $ which is the same as the Wilson line for $\phi$.  

\subsection{The Dirichlet E brane } 
We start in the closed string channel and consider the annulus diagram with UV and IR boundaries of size $\epsilon$ and $\Lambda$ . We insert the Dirichlet E brane states \eqref{D} at both ends and search for an open string description in terms of summing over local, cardy boundary conditions.  Mapping the annulus to the cylinder with length $l= \log \frac{\Lambda}{\epsilon}$ we have
  \begin{align}\label{annulus}
   \mathtikz{ \pairA{0cm}{0cm} \copairA{0cm}{0cm} \draw (0cm,-0.1cm) node {\footnotesize $e$}; \draw (0cm,1cm) node {\footnotesize $e$}} &= \braket{e|\tilde{q}^{\frac{1}{2}(L_{0}+\bar{L}_{0}-\frac{c}{12})}|e} = 2R \pi^2 \int_{0}^{2 \pi R} d \phi_{a}  \int_{0}^{2 \pi R}d \phi_{b}  \langle\langle 0|\phi_{a} \rangle \langle\phi_{b}  |0 \rangle\rangle   \braket{\phi_{a}|\tilde{q}^{(L_{0}-\frac{c}{24})}|\phi_{b} }\nn
    &= C \sqrt{\frac{ \pi }{ l }}   \int_{0}^{2 \pi R} d \phi_{a}  \int_{0}^{2 \pi R}d \phi_{b} \sum_{n \in \mathbb{Z}} e^{\frac{-1} {l} ( 2  \pi R n +  \phi_{b}-\phi_{a}  )^{2}}\, \eta(\tilde{q} )^{-1}\nn 
     &= C   \int_{0}^{2 \pi R} d \phi_{a}  \int_{0}^{2 \pi R}d \phi_{b} \sum_{n \in \mathbb{Z}} e^{\frac{-1 }{l} ( 2  \pi R n +  \phi_{b}-\phi_{a}  )^{2}}\, \eta(q )^{-1}\nn
 \tilde{q} &= \exp{-2l}, \quad  q= \exp{\frac{-2\pi^{2}}{l}  } \end{align}
where we noted that $C=  2R \pi^2  \langle\langle 0|\phi_{a} \rangle \langle\phi_{b}  |0 \rangle\rangle  $ is a constant independent of the boundary conditions, and $\eta(q)$ is the Dedekind eta function.  In the second equality, we have evaluated the amplitude and applied a Poisson resummation to the sum over momentum modes in \eqref{DN} .
In the final equality we used the transformation properties of $\eta$ under a modular transformation to obtain:
\begin{align}
    Z_{ab}(q) = \sum_{k \in \mathbb{Z}} e^{\frac{-1 }{l} ( 2  \pi R k +  \phi_{b}-\phi_{a}  )^{2}}\, \eta(q )^{-1}
\end{align}
By observing that $Z_{ab} = \tr_{ab} q^{L_{0}-\frac{c}{24}}$, we see immediately that this is a sum over BCFT partition functions with Dirichlet boundary condition.  However, let us work out the corresponding open string extension in detail, making direct use of the Euclidean path integral to compute the associated cobordisms.

If we define 
\begin{align}
    \phi_{\epsilon}&= \phi_{a}\nn
    \phi_{\Lambda}&= \phi_{b} + 2 \pi R k 
\end{align}
and note that $\phi_{\Lambda}$ ranges over $\mathbb{R}$ as $k$ ranges over $\mathbb{Z}$, then the annulus cobordism can be expressed as
\begin{align}\label{DM}
        \mathtikz{ \pairA{0cm}{0cm} \copairA{0cm}{0cm} \draw (0cm,-0.1cm) node {\footnotesize $e$}; \draw (0cm,1cm) node {\footnotesize $e$}}= C    \int_{0}^{2 \pi R} d \phi_{\epsilon} \int_{ -\infty}^{\infty} d \phi_{\Lambda}  e^{\frac{-1}{l} (  \phi_{\Lambda}-\phi_{\epsilon }  )^{2}}  \eta^{-1}(q)
\end{align}

Though not necessarily unique, there is a natural open string extension that reproduces this cobordism as the trace of a reduced density matrix.  This assigns an extended Hilbert space to an interval with E brane boundary labels $e$:
\begin{align}
I_{ee} \rightarrow \bigoplus_{\phi_{a},\phi_{b} \in [0, 2\pi R)} \mathcal{H}_{\phi_{a},\phi_{b}}
\end{align}
The states in each superselection  sector are obtained by quantizing
\begin{align}
\phi(x)&= \phi_{a} + \frac{\phi_{b}- \phi_{a} }{l} x +  2\pi n R \frac{x}{l} +\sum_{k>0} a_{k} \sin k x  \nn
\phi_{a},\phi_{b}  &\in [0,2 \pi R], \quad n \in \mathbb{Z}
\end{align} 
The corresponding boundary primaries are labelled by $\ket{\phi_{a},\phi_{b},n}$, where $n$ and $\frac{\phi_{b}- \phi_{a}}{2 \pi R} $ label the integer and fractional parts of the winding number of an open string.  We have separated these because  $\phi_{a},\phi_{b}$ as super selection labels that do not change under any allowed interactions- in string theory language these are location of D branes.  This means that the allowed density matrice will not connect sectors with different $\phi_{a},\phi_{b} $\footnote{In the BCFT language, each sector correspond to the insertion of a boundary condition changing operator of dimension \eqref{h} }.  On the other hand, a general density matrix can connect sectors with different integer winding numbers $n$, as we will see in the next subsection when we consider bulk winding states.

\begin{figure}[h]
\centering
\includegraphics[scale=.5]{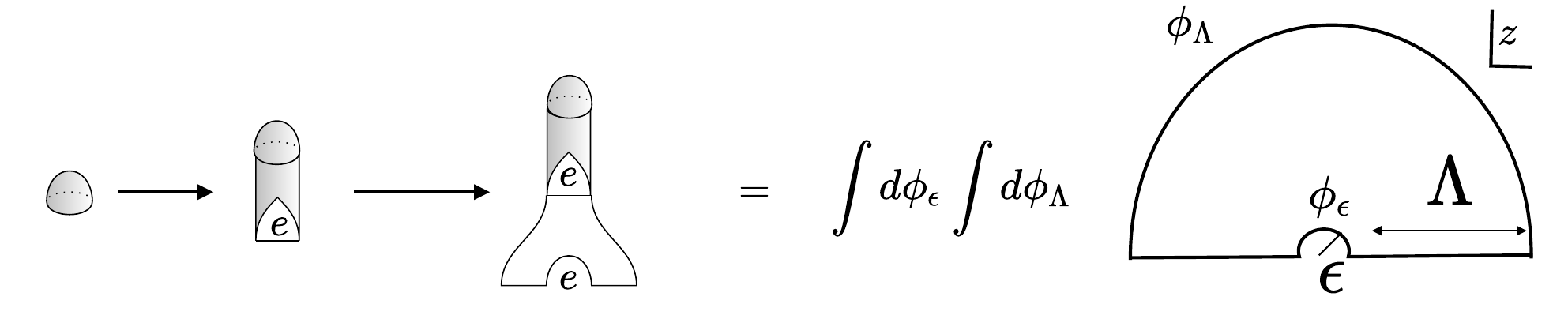}
\caption{ The factorization cobordism. In this conformal frame, the origin and the point at infinity are regulated by semi-circles of size  $\epsilon$ and 
$\Lambda$, where we integrate over boundary conditions $\phi_{\epsilon},\phi_{\Lambda}$. }
\label{Factor}
\end{figure}

The factorization cobordisms associated to this open string extension is:
\begin{align} \label{fcob}
    \mathtikz{ \zipper{0cm}{0cm}\draw (.0cm,-.7cm) node {\footnotesize $e$};}: \mathcal{H}_{S^1} &\rightarrow \bigoplus_{\phi_{0} \in [0,2\pi R)} \mathcal{H}_{\phi_{0}\phi_{0}}\\
    \mathtikz{ \deltaA{0cm}{0cm} 
\draw (-0.8cm,-0.5cm) node {\footnotesize $a$};
\draw (0cm,-1 cm) node {\footnotesize $e$};
\draw (0.8cm,-0.5cm) node {\footnotesize $c$};}: \mathcal{H}_{ac} &\rightarrow  \bigoplus_{ \phi_{0} \in [0,2\pi R)} \mathcal{H}_{a \phi_{0} } \otimes \mathcal{H}_{\phi_{0} c}
\label{OP}
\end{align}
Applying these to the closed string vacuum on the circle gives the factorized vacuum state in  figure \ref{Factor}.   In the interval splitting cobordism, we choose a conformal frame in which the states in the tensor product live on the left and right half of the real axis, which differs from the standard frame for the interaction cobordism.  This choice will facilitate comparison with the closed string amplitude \eqref{annulus} as well as the usual entanglement calculations.  

Let us evaluate this vacuum factorization cobordism explicitly by computing the Euclidean path integral, and show that it is consistent with \eqref{DM}. Since we have a free theory, for each sector this amounts to evaluating the on shell action with given field configurations $\phi_{R} , \phi_{L}$ on the two half lines.   
For this purpose, it will be convenient to make a change of variables to a strip geometry with coordinate $w =\log z= x +i t $. 
Defining  $l= \log \frac{\Lambda}{\epsilon},\quad  k= \frac{ \pi n}{l},  n \in \mathbb{Z}$, we expand the field configurations as
\begin{align}
\phi_{R}(x)&= \phi_{\epsilon} + \frac{\phi_{\Lambda}-\phi_{\epsilon}}{l} (x-\log\epsilon) +\sum_{k>0} R_{k} \sin k (x-\log \epsilon) \nn
\phi_{L}(x)&=  \phi_{\epsilon} +\frac{\phi_{\Lambda}-\phi_{\epsilon}}{l}  (x-\log \epsilon)+ \sum_{k>0} L_{k} \sin k (x-\log \epsilon)
\nn
\phi_{\epsilon} &\in [0,2 \pi R)\quad
\phi_{\Lambda} -\phi_{\epsilon} \in \mathbb{R}
\end{align}

The on shell action for the configuration with these boundary configurations is
\begin{align} 
\frac{1}{2\pi } \int_{\pd M } \phi \pd_{n} \phi = \exp \left(- \frac{ (\phi_{\Lambda}-\phi_{\epsilon})^2}{ 2  l } -S[L,R] \right) , 
\end{align}  
where $S[L,R]$ is the contribution from the oscillating modes.   This gives the factorized state 
\begin{align} \label{Dstate}
    \ket{\Psi}_{D}&=   \bigoplus_{\phi_{\epsilon}=0}^{2 \pi R}   \bigoplus_{\phi_{\Lambda}=-\infty}^{\infty}\left(  e^{- \frac{\pi}{2l} (\phi_{\Lambda}-\phi_{\epsilon})^{2}}  \sqrt{N} \int D[R(x)] D[L(x)]  e^{-S[L,R]}   \ket{ \phi_{\Lambda},\phi_{\epsilon}, L(x)}\otimes \ket{ \phi_{\epsilon}, \phi_{\Lambda}, R(x)}\right)
 \end{align} 
where  $N$ is a normalization constant from the fluctuation determinant of the oscillators.  On this state, the reduced density matrix on one interval coincides with the density matrix of equation \eqref{DM}, up to a constant.  Note that in it's factorized form, it is the entanglement of the labels $\phi_{\epsilon}, \phi_{\Lambda}$ that identify this state as living the circle rather than an interval.  

While we have emphasized above that the integer winding numbers $n$ are not super selection labels,  the reduced density matrix of the \emph{vacuum} is nevertheless block diagonal in $n$. As a result, it is also block diagonal in in $\phi_{\epsilon}, \phi_{\Lambda}$, which are just the  boundary conditions of the open string.  For this reason the entanglement entropy will again take the form of \eqref{eq:EE}, with $\phi_{\epsilon}, \phi_{\Lambda}$ playing the role of the label $k$.  In particular, we see that the probability factor $P_{k}$ is  determined by the zero mode part of the on shell action \cite{Lin:2018bud}:  
\begin{align}
P_{\phi_{\epsilon}\phi_{\Lambda}} &= \frac{1}{Z_{0}} \exp ( - \frac{1}{2\pi }\int_{\pd M } \phi \pd_{n} \phi)  = \frac{1}{Z_{0}}\exp (- \frac{ (\phi_{\Lambda}-\phi_{\epsilon})^2}{   l })\nn
 Z_{0} &=  2 \pi R \sqrt{\pi l}
\end{align}  
which gives the ``edge mode" entropy
\begin{align}
S_{\text{edge}} = -\sum_{\phi_{\epsilon},\phi_{\Lambda} }P_{\phi_{\epsilon}\phi_{\Lambda}} \log P_{\phi_{\epsilon}\phi_{\Lambda}} = \log 2 \pi R +  \frac{1}{2}(1+\log l+ \log \pi ) 
\end{align} 
The first term is due to the equal mixture of states with different $\phi_{\epsilon}$,  and the second term is from the gaussian probablity factor for  $\Delta \phi =\phi_{\Lambda}-\phi_{\epsilon}$.    The $ \frac{1}{2} \log l$ factor is crucial as it cancels against a $- \frac{1}{2} \log l$ from the entropy due to the oscillators:
 \begin{align}
     S_{\text{osc}} &=( 1- n\pd_{n}) \log \eta^{-1}( q^{n } ) =-\frac{1}{2}(1+ \log l - \log \pi ),  
     \end{align}
   where we applied a modular transformation and the large $l$ limit: 
   \begin{align}
     \eta(q^{n}) &= \sqrt{\frac{n\pi}{l}} \eta(\tilde{q}^{\frac{1}{n}})\rightarrow  \sqrt{\frac{n\pi}{l}} \exp( \frac{l}{12n}) 
 \end{align}
The total entropy is then 
 \begin{align}\label{DEE}
 S= \frac{l}{6}   +\log{ 2 \pi R}  +\log \pi + O (e^{-l}) 
 \end{align}    



\subsection{The Neumann E Brane}
Here we write the  Neumann E brane as in equation \eqref{N}, which can be obtained from the Dirichlet E brane by T-duality: 
\begin{align}
R\rightarrow \frac{1}{2R} 
\end{align} 
The cylinder amplitude with Neumann E brane boundary states inserted is
 \begin{align}\label{NE}
   \mathtikz{ \pairA{0cm}{0cm} \copairA{0cm}{0cm} \draw (0cm,-0.1cm) node {\footnotesize $e$}; \draw (0cm,1cm) node {\footnotesize $e$}} 
 &= C'   \int_{0}^{\pi/ R} d \tilde{\phi}_{a}  \int_{0}^{ \pi/ R}d \tilde{\phi}_{b} \sum_{n \in \mathbb{Z}} e^{\frac{-1 }{l} (  \frac{\pi n}{R} +  \tilde{\phi}_{b}-\tilde{\phi}_{a}  )^{2} }\, \eta(q )^{-1}\nn
&= \tr   \bigoplus_{\tilde{\phi}_{\epsilon} =0}^{ \pi/ R}  \bigoplus_{ \tilde{\phi}_{\Lambda}=-\infty}^{\infty}  C' e^{\frac{-1 }{l} (  \tilde{\phi}_{\Lambda}-\tilde{\phi}_{\epsilon}  )^{2}}  \exp(- 2 \pi H_{\text{osc}} )\nn
\tilde{\phi}_{\epsilon}&=\tilde{\phi}_{a} ,\quad \tilde{\phi}_{\Lambda}=\tilde{\phi}_{b} + \pi n/R  \in \mathbb{R}
 \end{align}
 where $H_{\text{osc}}$ refers the the oscillator part of the Hamiltonian whose trace gives $\eta(q)^{-1}$.
 An open string extension consistent with this density matrix satisfies Neumann boundary conditions $ \pd_{x}\phi=0$ and couples to background ``Wilson line" $\tilde{\phi}_{a}, \tilde{\phi}_{b} \in [0, \frac{\pi}{R}]$, which are the $T$ dual superselection labels. On the cylinder, these correspond to adding boundary terms in the action at the two circular boundaries\footnote{Since this boundary term itself is a total derivative, it doesnt alter the equation of motion or boundary condition $\pd_{n} \phi =0$. However it does change the symplectic structure}: 
\begin{align}
    S_{\pd} =  \frac{\tilde{\phi}_{b}}{\pi} \oint_{x=\log \Lambda }  dt\,  \pd_{t}\phi -\frac{\tilde{\phi}_{a} }{\pi}\oint_{x=\log \epsilon}  dt\,   \pd_{t}\phi 
\end{align}

For Neumann boundary conditions, $\phi(x,t)$ can be expanded as
\begin{align}
\phi(x,t) &= \phi_{0} + p t + \sum_{n \neq 0 } a_{n}(t) \cos(k (x-\log\epsilon)) 
\end{align}
while the back ground Wilson lines change the canonical momentum to 
\begin{align}
   \Pi(x,t)  =  \frac{1}{\pi} \left( \pd_{t}\phi +\tilde{\phi}_{b} \delta (x- \log \Lambda )- \tilde{\phi_{a}} \delta (x- \log \epsilon)  \right)
\end{align}
We define the zero mode of $\Pi$ by
\begin{align}
    \Pi_{0}&=  \frac{1}{l} \int \Pi (x,t)  dx =\frac{1}{\pi} \left(p + \frac{\tilde{\phi}_{b}-\tilde{\phi}_{a} }{l}  \right)  
\end{align} and impose the canonical commutation relations
\begin{align}
    [\Pi(x), \phi(y)]&= i \delta(x-y) \nn
  [\Pi_{0}l,\phi_{0}]= i,  
\end{align}
where the commutator for the zero modes come from integrating the commutator for the fields. 

Since $\Pi_{0} l $ is the canonically conjugate variable to the circle variable $\phi_{0} \in [ 0,2 \pi R] $, it must be quantized in units of $\frac{1}{R}$ so that the wavefunction $e^{ i \Pi_{0} \phi_{0} l } $ is single valued.  Writing the physical momentum $p$ in terms of the quantized eigenvalues of $\Pi_{0}$ then gives 
\begin{align}
    p &= \frac{1}{l}  \left( \frac{n \pi}{R} +(\tilde{\phi}_{b}-\tilde{\phi}_{a}) \right)\quad n\in \mathbb{Z}\nn &=\frac{1}{l}(\tilde{\phi}_{\Lambda}-\tilde{\phi}_{\epsilon})   \nn
    \tilde{\phi}_{\Lambda}&\equiv \frac{\pi n}{R} + \tilde{\phi}_{b}\in \mathbb{R},\quad \tilde{\phi}_{\epsilon}\equiv \tilde{\phi}_{a}\in [0,\pi/R],
\end{align}
showing that the physical momentum can have a fractional contribution from the Wilson lines.  
The Hilbert space of an interval can then be described in terms of modes of $\Pi$, with the zero mode $\tilde{\phi}_{a},\tilde{\phi}_{b}$ playing the role of super selection labels :
\begin{align}
I_{ee} \rightarrow \bigoplus_{\tilde{\phi}_{a},\tilde{\phi}_{b} \in [0,\frac{\pi}{R}]} \mathcal{H}_{\tilde{\phi}_{a},\tilde{\phi}_{b}}
\end{align}
The gluing rules are given as in \eqref{fcob}, with the Wilson line variables replacing the Dirichlet boundary conditions.  

Given this choice of the interval Hilbert space, we can again compute the factorized ground state directly from the path integral. In this case, evaluating the on shell action $\frac{1}{2} \int_{\pd M} \phi \pd_{n} \phi $ seems to give zero because $\pd_{n}\phi = \pd_{x} \phi =0 $ along the two boundary circles of the cylinder. But we must not forget that the $\phi(x,t)$ is multivalued along the $t$ direction , so there is a branch cut at $t=2 \pi $ that acts effectively as a boundary.  The on shell action comes from this cut and gives
\begin{align}
   \frac{1}{2\pi} \int_{\text{cut} }\phi \pd_{t} \phi  dx =  l  p^{2} =\frac{1}{l} (\tilde{\phi}_{\Lambda} - \tilde{\phi}_{\epsilon} )^{2} 
\end{align}
leading to  the factorized vacuum state 
\begin{align}
    \ket{\Psi}_{N}&=   \bigoplus_{\tilde{\phi}_{\epsilon}=0}^{ \pi/ R}   \bigoplus_{\tilde{\phi}_{\Lambda}=-\infty}^{\infty}\left(  e^{- \frac{1}{2l} (\tilde{\phi}_{\Lambda}-\tilde{\phi}_{\epsilon})^{2}}  \sqrt{N} \int D[R(x)] D[L(x)]  e^{-S[L,R]}   \ket{ \tilde{\phi}_{\Lambda},\tilde{\phi}_{\epsilon}, L(x)}\otimes \ket{ \tilde{\phi}_{\epsilon}, \tilde{\phi}_{\Lambda}, R(x)}\right)
 \end{align} 
Taking the norm gives back the trace of the Neumann density matrix in \eqref{NE}
\begin{align}
    \braket{\Psi_{N}|\Psi_{N}} 
  = \int_{0}^{\pi/R} d\tilde{\phi}_{\epsilon} \int_{-\infty}^{\infty} d w_{\Lambda } \exp( - \frac{1}{ l}(  \tilde{\phi}_{\Lambda}-\tilde{\phi}_{\epsilon} )^{2} ) \eta^{-1}(q) 
    \end{align}
    
Just like the state, the entropy is related to the Dirichlet case by T-duality: 
 \begin{align}\label{NEE}
 S= \frac{l}{6}   +\log{  \frac{\pi}{ R}}  +\log \pi + O (e^{-l})
 \end{align}    
\subsection{Comments on edge mode EE and shift symmetry }
In the previous sections, we saw that the sum over the ``D brane" edges modes $\phi_{a,b}$ or $\tilde{\phi}_{a,b} $ led to a subleading constant term in the entanglement entropy for the vacuum state. This is due to the shift symmetry of the vacuum under
\begin{align}
     \phi \rightarrow \phi + a \quad , a \in [0,2 \pi R]\nn
      \tilde{\phi} \rightarrow  \tilde{\phi}+ \tilde{a} ,  \quad \tilde{a} \in [0, \pi/ R]
\end{align}
As noted earlier, the Neumann/Dirichlet E brane manifestly preserves these symmetries: picking a single Neumann or Dirchlet boundary condition breaks one of these symmetries, but this is restored by summing over boundary conditions when we glue together intervals.  More abstractly, the gluing is a projection onto a subspace of the tensor product which is invariant under the simultaneous shift of either $\phi$ or $\tilde{\phi}$.   In this way, this gluing is reminiscent of the entangling product in gauge theories \cite{Donnelly:2016auv}, with the shift symmetry playing a similar role as the boundary symmetry.

As a consequence, the reduced density matrix must also be shift invariant, which is why the factorized wavefunction on only depends on  $\phi_{b}-\phi_{a}$, or $\tilde{\phi}_{b}-\tilde{\phi}_{a}$. 
Thus, the shift symmetry leads to an edge mode degeneracy e.g. of  ``size" $2 \pi R $ for the Dirichlet extension. This  explains extra entanglement entropy of $S=\log 2 \pi R $. Because the superselection labels are continuous, a choice of measure has to be made to sum over them. We have implicitly chosen such a measure in the integration over boundary conditions. In particular, for the degenerate Dirichlet edge mode $\phi_{a}$ we have
\begin{align}
    \int_{0}^{2 \pi R}  d \phi_{a}=  \int_{0}^{1}  (2 \pi R ) d \phi_{a}
\end{align}
so we treat $2\pi R $ as the chosen measure.   This is reminiscent of gauge theory and JT gravity \cite{Jafferis:2019wkd} \cite{Lin:2018xkj}, where the same $\log (\text{measure})$ factor appears as an edge mode contribution to the entanglement entropy.   Notice also that the entanglement entropy is not invariant under a change of measure. Given normalized probability factors $P(\alpha)$, a change in measure rescales the probablities according to: 
\begin{align}
   1= \int d\alpha P(\alpha) &=\int c\, d \alpha P'(\alpha)\nn P'(\alpha) &= \frac{P(\alpha)}{c}
\end{align}
The edge mode entropy then changes by a constant:
\begin{align}
     S_{\text{edge}}'= \int c \,  d \alpha \,  P'(\alpha ) \log P'(\alpha) = S_{\text{edge}}  - \log c 
\end{align}
Thus the difference in the EE for Neumann and the Dirichlet E brane can be attributed to the different measures on the superselection labels. This also accounts for small discrepancy with the edge entropy computed in \cite{Michel:2016fex}

To get a hint of how things work for more general CFT's, we can interpret the effect of our factorization on the modes of the $U(1)$ symmetry algebra of the boson.   On an interval, the zero mode $J_{0}$  is either the momentum or the winding number, depending on whether we impose Neumann or Dirichlet boundary conditions \footnote{ On the circle, both the momentum and the winding number belong to the operator algebra.  However the choice of Neumann or Dirichlet boundary conditions on the interval removes either the winding number or the momentum from the local subalgebra.  The conjugate variable  $\tilde{\phi}$ or $\phi$ thus becomes the center of the subalgebra on the interval.  The eigenvalues of the center then become super-selection labels, because we have removed the conjugate operator that could change them \cite{Lin:2018bud}. } .   In either case, the gluing rules imply a matching of the zero mode on either side of the entangling surface.  The action of $J_{0}$ on the global vacuum state thus factorizes as : 
\begin{align}
    \Delta (J_{0}) = 1 \otimes J_{0}+ J_{0}\otimes 1
\end{align}
where the $J_{0}$ on the LHS denotes the charge on the total space.  We can think of $\Delta$ as giving a representation of the global charge on the tensor product Hilbert space. This naturally leads to the question of how $\Delta$ acts on the local charges $J_{n}$ for $n \neq 0$: we will elaborate on this in section 4.  Before doing so, we first turn the factorization of the primaries of the compact boson, corresponding to the winding and momentum states on the circle.

\subsection{Factorization of primaries states in terms of three point functions}
\begin{figure}[h]
\centering
\includegraphics[scale=.3]{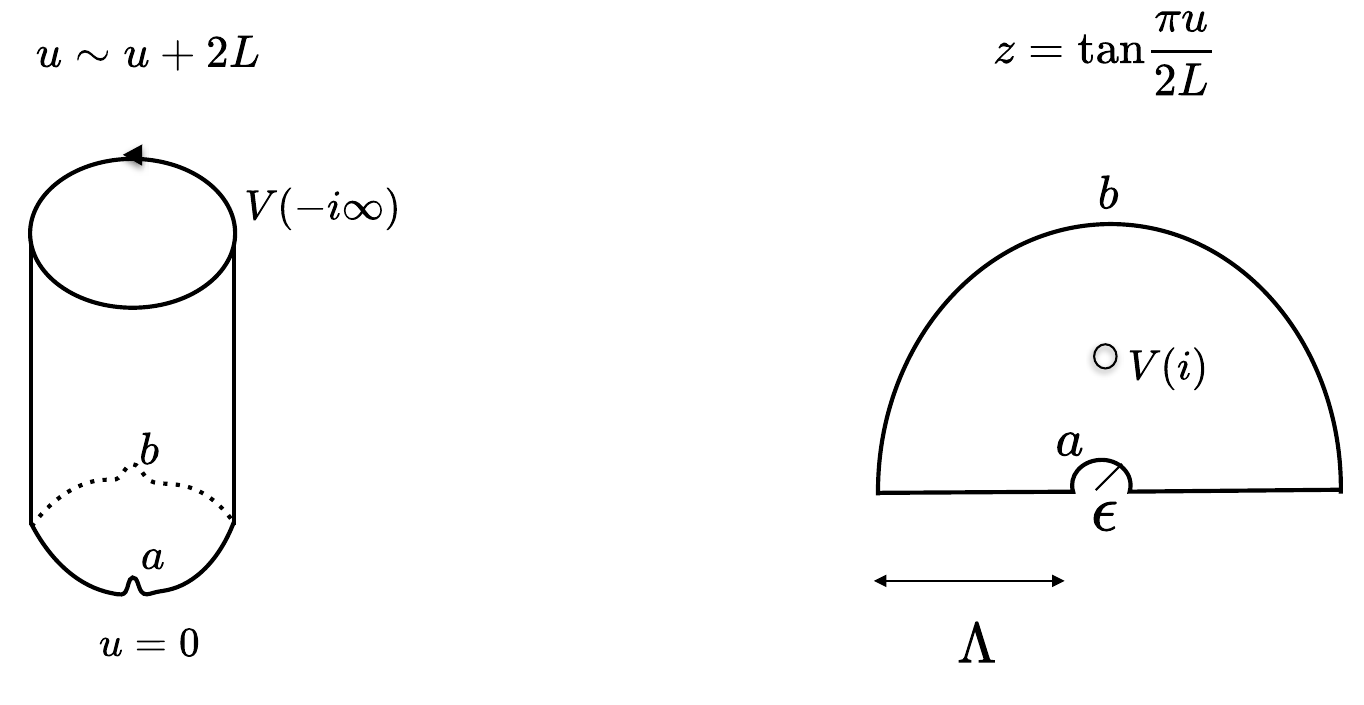}
\caption{Preparation of the state and the factorization map is combined by the Euclidean path integral on a half infinite cylinder, with regulator semi circles around the entangling surface. Our factorization map corresponds to summing over Cardy boundary conditions at the entangling surface.}
\label{circle}
\end{figure}

\begin{figure}[h]
\centering
\includegraphics[scale=.35]{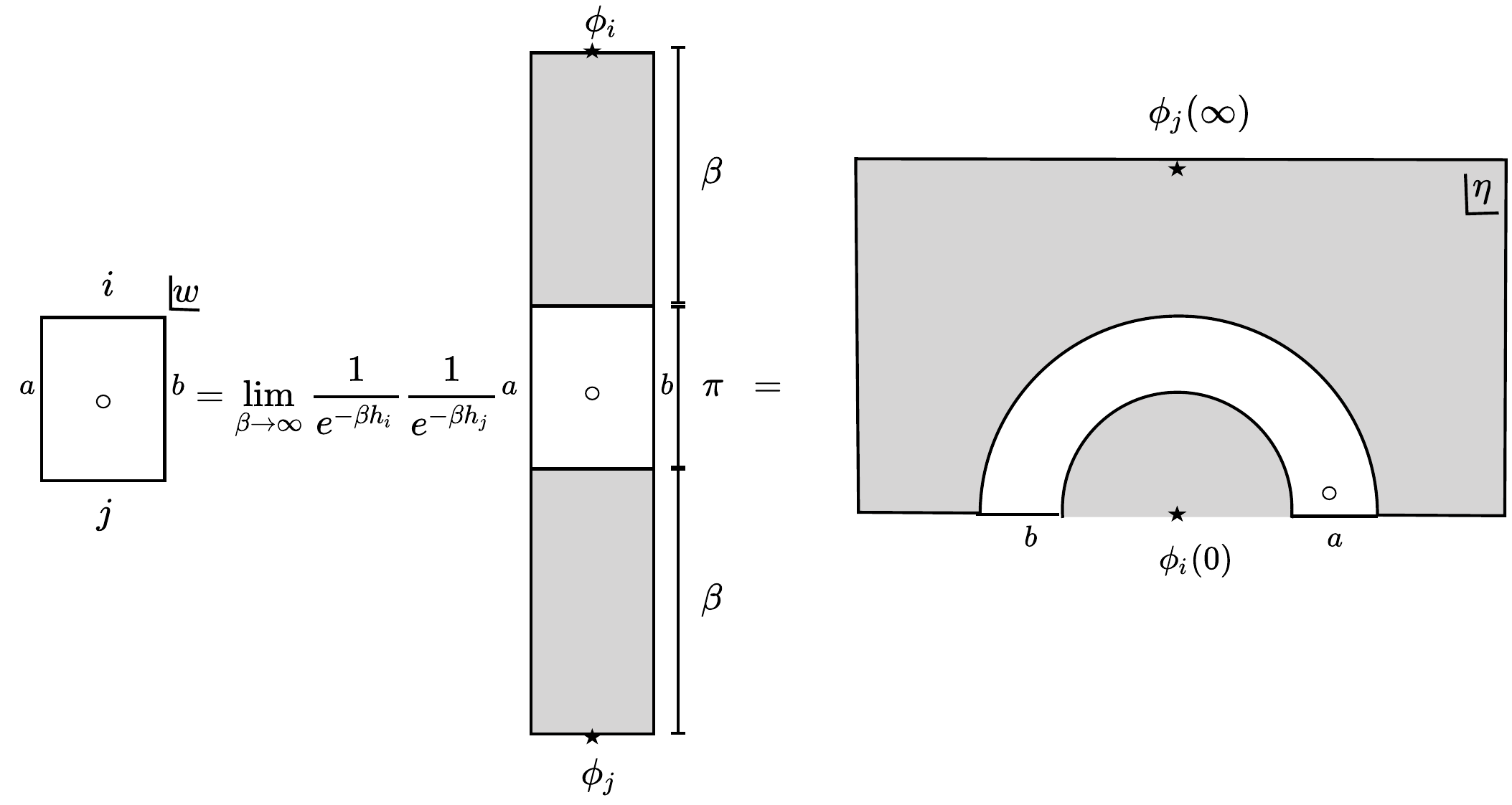}
\caption{The wavefunction for a factorized state can be expressed as a correlation function in the upper half $\eta$ plane as in the right figure. The circle represents the insertion of a bulk operator which determines the global state. In the middle figure, the shaded regions describe the state preparation along a half infinite strip, properly normalized by dividing by $e^{-\beta h_{i} }$.  These factors are cancelled when transformed in to the $\eta$ plane on the right.  The un-shaded region corresponds to modular flow, which rescales the \emph{normalized} primary state $\phi_{i}(0) \ket{0}$  by a factor of $e^{-\frac{\pi}{2 l}h_{i}}$.}
\label{TFD2p}
\end{figure}

In order to discuss the factorization of the winding and momentum states, it is useful to express the factorization cobordism directly in terms of the CFT data.   In general a state $\ket{V}$ on a spatial circle of length $2L$ can be prepared by Euclidean evolution along a half infinite cylinder geometry with prescribed boundary condition at infinity (figure \eqref{circle}).
 To compute the factorization map we cut small semi-disks out of the cylindrical geometry which separates the circle into two intervals. This is conformally equivalent to  the \emph{regulated} $z$  half plane in the right of figure \eqref{circle} via the mapping 
\begin{align} 
z= \tan\frac{\pi u}{2 L} ,
\end{align} 
where a vertex operator $V$ has been at $z=i$ to specify the state.  As we showed in the previous section, choosing an E brane boundary condition for the boson means we can factorize the state as 
\begin{align} \label{V}
\ket{V}&= \sum_{i,j,a,b} \braket{i\,a\,b,j\,b\,a|V} \ket{i\,a\,b}\otimes \ket{j\,b\,a} \
\end{align} 
where $\braket{i\,a\,b,j\,b\,a|V}$ is the path integral on this geometry with either Neumann or Dirichlet boundary conditions $a,b$.
As in the case of the vacuum, we make a further mapping to the finite strip geometry via $w=\log \frac{z}{\epsilon}$, which gives the left diagram of figure \ref{TFD2p}. Noticed that this conformal mapping implements a CPT transformation on the state living on the left interval.  In this context it is best interpreted as a state-channel duality that takes a tensor product state into a linear map.  This maps the wavefunction of the factorized state to an amplitude
\begin{align}
\braket{i \,a \,b,j\, b\, a|V} &\rightarrow   \braket{i, a, b| ^{-\frac{\pi^{2} }{2 l } L_{0} } V e^{-\frac{\pi^{2}}{2 l }L_{0} }|j,a,b }\\
L_{0} &=  \frac{1}{2\pi i} \oint\eta T (\eta) d \eta,  \quad \eta = \exp( \frac{i \pi w}{l} ) 
\end{align}
where $a,b$ labels physical boundary conditions, and $\ket{i,a,b}, \ket{j,a,b}$ are normalized boundary states that diagonalize the modular Hamiltonian $L_{0}$, expressed above as a dilatation on the upper half $\eta$ plane. The arrow refers to the state channel duality, which reverses the order of the boundary labels \footnote{We have imposed the matching of the boundary labels in accordance with the OPE structure of the BCFT. This enforces the superselection rule that no operators can change the boundary labels}.   
As shown in the middle diagram in figure \ref{TFD2p}, the operator state correspondence maps these normalized states to half infinite strips with operator inserted at infinity, divided by a normalization factor.  This allows us to write the factorized wavefunction as a CFT three point function most conveniently evaluated in the $\eta$ plane.   Denoting the \emph{normalized} boundary primaries as $V^{ab}_{i},V^{ba}_{j} $ we have
\begin{align}\label{3pt}
    \braket{i\,a\,b,j\,b\,a|V} &\rightarrow    e^{-\frac{\pi }{2l }(h^{ab}_{i}+h^{ab}_{j}) } \braket{V_{i}^{ab}(0) \,V(\eta)\, V_{j}^{ba} (\infty)  } \nn
    \eta&= \exp(-\frac{\pi^2}{2 l }) \exp (\frac{i \pi \log \epsilon}{l})
\end{align}
For large $l$, the bulk insertion point $\eta$ is close to the boundary and we can apply the bulk to boundary OPE's.   We find that to leading order
\begin{align}
    \braket{i\,a\,b,j\,b\,a|V} &\rightarrow  \sum_{k} (2 \text{Im} \eta )^{h_{i}-h_{V}-\bar{h}_{V}}  e^{-\frac{\pi }{2l }(h^{ab}_{i}+h^{ab}_{j}) } C_{Vk}^{b} C^{bab}_{kij} \nn
    &\sim \sum_{k} (\frac{2 \pi \log \epsilon}{l}  )^{h_{i}-h_{V}-\bar{h}_{V}}  e^{-\frac{\pi }{2l }(h^{ab}_{i}+h^{ab}_{j}) } C_{Vk}^{b} C^{bab}_{kij}
\end{align}

As a simple check, we can apply \eqref{3pt} to  the vacuum state for which there is no bulk insertion\footnote{We make the usual abuse of notation in which the insertion at infinity is understood to be accompanied with a conformal factor $\eta^{2 h_{i}} $ as $\eta \rightarrow \infty$}. This gives
\begin{align}
      \braket{i\,a\,b,j\,b\,a|V} &\rightarrow e^{-\frac{\pi^{2}}{2l }(h^{ab}_{i}+h^{ab}_{j}) } \braket{  V_{i}^{ab}(0)  V_{j}^{ba}(\infty) }=e^{\frac{-\pi^{2} }{l} h^{ab}_{i} } \delta_{ij} \nn
      \ket{V} &= \sum_{i, a,b} e^{\frac{-\pi^{2} }{l} h^{ab}_{i} } \ket{i,a,b} \ket{i,b,a} 
\end{align}
 This gives a generalized version of the usual thermofield double state which describes the Minkowski vacuum.  For the compact boson  with Dirichlet-Dirichlet boundaries, this gives  exactly the factorization amplitude of \eqref{Dstate} when we insert the conformal dimension
 \begin{align}\label{h}
     h^{i}_{ab} = \frac{1}{2}( \frac{\phi_{b}-\phi_{a}}{\pi} +2 R n_{i} )^{2}
 \end{align}
where $n_{i}$ is the integral part of the winding number.

Now let us consider the factorization of nontrivial winding and momentum states for the free boson 
\begin{align} 
\ket{k,\tilde{k} }&= V_{k,\tilde{k} }\ket{0} \nn
V_{k,\tilde{k} }&=:\exp(i \tilde{k} \tilde{\Phi} )\exp(i k \Phi ) :
\end{align} 
where $k=\frac{n}{R}$ and $\tilde{k}=\frac{\tilde{n}}{\tilde{R}}$ labels the momentum and the winding, notated as the T-dual momentum.

For simplicity lets first restrict to states with zero winding, $\tilde{k}=0$.  Inserting the Neumann E brane means we should factorize this state into an equal superposition of boundary states with Neumann-Neumann boundary conditions.  For general Wilson lines, the boundary states can be expressed in terms of boundary operators: \footnote{The boundary vertex operators undergo a boundary normal ordering that differs from the bulk normal ordering, but we keep the same notation to avoid clutter} 
\begin{align}
\ket{i\, \tilde{\phi}_{a}\,\tilde{\phi}_{b}}&= \exp\left(i \frac{n_{i}}{R} \Phi(0)\right) \ket{ \tilde{\phi}_{a},\tilde{\phi}_{b}}=\exp\left(i k^{ab}_{i} \Phi(0)\right) \ket{0,0} \nn
k^{ab}_{i}&= \frac{n_{i}}{R} + \frac{\tilde{\phi}_{b} -\tilde{\phi}_{a}}{\pi}  \nn
\tilde{\phi}_{a},\tilde{\phi}_{b} &\in [0,\frac{\pi}{R}]
\end{align}
where $k^{ab}_{i}$ labels the total open string momentum on an interval, which includes both the integer and fractional part.   The factorized wave function is then determined by the three point function. 
\begin{align}\label{NN}
       \braket{i\,\tilde{\phi}_{a}\,\tilde{\phi}_{b},j \,\tilde{\phi}_{b}\,\tilde{\phi}_{a}|k,\tilde{k}=0} &\rightarrow    e^{-\frac{\pi^{2} }{2 l }(h^{ab}_{i}+h^{ba}_{j}) } \braket{ 
       \exp\left(-i k^{ab}_{i} \Phi(0)\right) \exp\left(i k \Phi(\eta)\right)\exp\left(-i k^{ba}_{j} \Phi(\infty)\right)
         }\nn
        & =e^{-\frac{\pi^{2} }{2 l }(h^{ab}_{i}+h^{ab}_{j}) } \delta(k-k_{i}-k_{j}) |\eta \bar{\eta} |^{\frac{k k^{ba}_{j}}{2}}|2 \text{Im} \eta|^{\frac{k^2}{4}} ,
\end{align}
where the conformal dimensions are $h^{ab}_{i} = \frac{1}{2}( \frac{\tilde{\phi}_{b}-\tilde{\phi_{a}}}{\pi} +  \frac{ n_{i}}{R} )^{2} $.

We can also factorize the same bulk state using the Dirichlet E branes.
The wavefunction then factorizes into boundary primaries with Dirichlet-Dirchlet boundary conditions $\phi_{a},\phi_{b}$
\begin{align}
    \braket{i\,\phi_{a}\,\phi_{b},j \,\phi_{b}\,\phi_{a}|k,\tilde{k}=0} &\rightarrow    e^{-\frac{\pi^{2} }{2l }(h^{ab}_{i}+h^{ba}_{j}) } \braket{ \exp\left(-i \tilde{k}^{ab}_{i} \Phi(0)\right)|\exp \left( i k \Phi(\eta) \right) | \exp\left(-i \tilde{k}^{ba}_{j} \Phi(0)\right)}\nn
    &= e^{-\frac{\pi^{2} }{2l }(h^{ab}_{i}+h^{ba}_{j}) } \delta ( \tilde{k}^{ab}_{i}+\tilde{k}^{ba}_{j})  \frac{\exp(i k \phi_{a} )}{| \eta - \bar{\eta}|^{\frac{k^{2}}{4} }} (\frac{\eta}{\bar{\eta}})^{\frac{k(\phi_{b}-\phi_{a})}{2 \pi i}}
\end{align}
where we defined $\tilde{k}^{ab}_{i}= \frac{\tilde{n}_{i}}{\tilde{R}} + \frac{\phi_{b} -\phi_{a}}{\pi}$.
\footnote{
This wavefunction can be computed from the one point function
\begin{align}
 \braket{ :\exp ( i \int J \Phi ) : } &=  : \exp(-\frac{1}{2} \int dz\,dw\, J(z)G_{D}(z,w)J(w)): \nn
 J &= \frac{n}{R} \delta (z- \eta) 
\end{align} 
where $G_{D}(z,w))$ is the greens function with Dirichlet boundary condition $\phi_{a}=\phi_{b}=0 $. For general $\phi_{a,b}$
we simply shift the field $\Phi$ in the above by a classical homogeneous solution with desired boundary conditions: 
\begin{align}
    \Phi \rightarrow \Phi + \frac{\phi_{b}-\phi_{a}}{2 \pi i} 
    \log \frac{\eta }{\bar{\eta}}
\end{align}  }

Taking the T-dual of this result gives the one point function of a winding operator with general Neumann boundary conditions.  This in turn can be combined with \eqref{NN} to obtain the factorization of a general state with $k$ and $\tilde{k}$ nonzero: 
 
\begin{align}
       \braket{i\,\tilde{\phi}_{a}\,\tilde{\phi}_{b},j \,\tilde{\phi}_{b}\,\tilde{\phi}_{a}|k,\tilde{k}} &\rightarrow \braket{i\,\tilde{\phi}_{a}\,\tilde{\phi}_{b}| \exp\left(i \tilde{k} \tilde{\Phi}(\eta)\right)\exp\left(i k \Phi(\eta)\right)|j \,\tilde{\phi}_{b}\,\tilde{\phi}_{a}}
         \nn
         &=\braket{i\,\tilde{\phi}_{a}\,\tilde{\phi}_{b}| \exp\left(i \tilde{k} \tilde{\Phi}(\eta)\right)|j \,\tilde{\phi}_{b}\,\tilde{\phi}_{a}}
        \braket{i\,\tilde{\phi}_{a}\,\tilde{\phi}_{b}| \exp\left(i k \Phi(\eta)\right)|j \,\tilde{\phi}_{b}\,\tilde{\phi}_{a}} \nn
        & =e^{-\frac{\pi^{2} }{2 l }(h^{ab}_{i}+h^{ab}_{j}) } \delta(k-k_{i}-k_{j}) |\eta \bar{\eta} |^{\frac{k k^{ba}_{j}}{2}}|2 \text{Im} \eta|^{\frac{k^2}{4}}\frac{\exp(i \tilde{k} \tilde\phi_{a} )}{| \eta - \bar{\eta}|^{\frac{k^{2}}{4} }} (\frac{\eta}{\bar{\eta}})^{\frac{k(\phi_{b}-\phi_{a})}{2 \pi i}}
\end{align}

\section{Co-product, Factorization and Fusion rules} In the previous section we applied the factorization cobordism to the primary state of the compact boson on a spatial circle.  Here we would like to generalize the factorization to descendants and for more general CFT's.  The key observation is that in each superselection sector the open string factorization cobordism \eqref{OP} can be unfolded into a chiral vertex operator which defines the fusion rules in a CFT \cite{Moore:1988qv}. 
\begin{figure}[h]
\centering
\includegraphics[scale=.35]{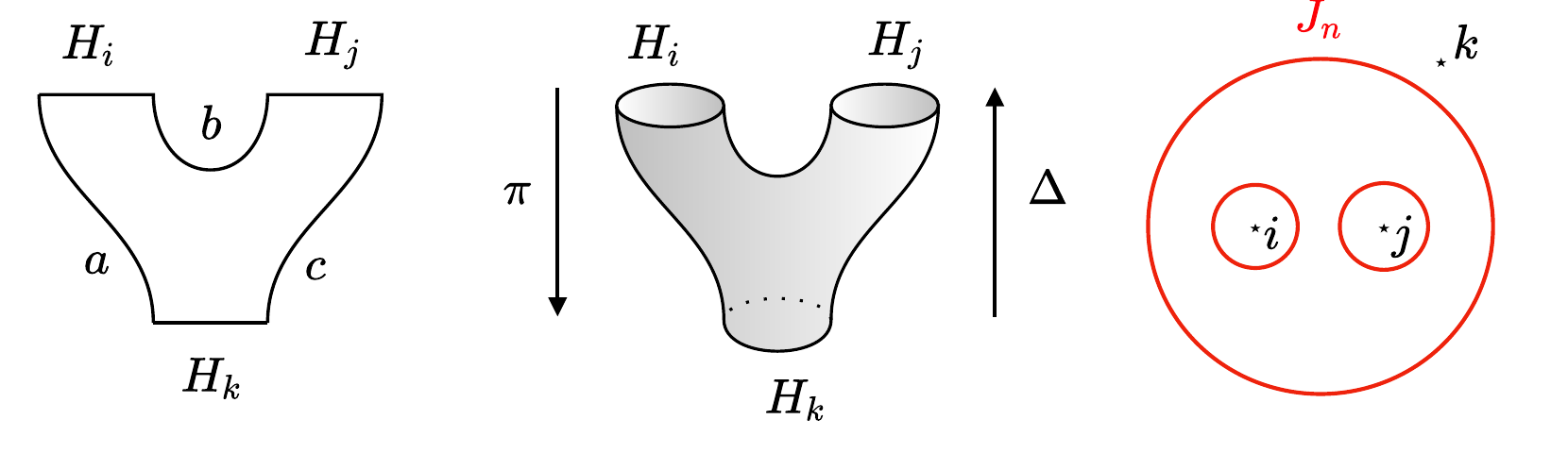}
\caption{The thrice punctured sphere defines a chiral vertex operator, which acts as an intertwining operator between the representations of the chiral algebra }
\label{cp3}
\end{figure} 
This is a path integral on a pair of pants geometry in figure \eqref{cp3}, which gives a linear map between irreps of the chiral algebra: 
\begin{align}\label{pi}
    \pi : \mathcal{H}^{ab}_{i} \otimes \mathcal{H}^{bc}_{j} \rightarrow \mathcal{H}^{ac}_{k}
\end{align} 
Moreover this map commutes with contour deformation, which allows the  charges $J_{n}=\oint dw \,w^n J(w) $ on  $\mathcal{H}^{ac}_{k}$ to be expressed in terms of charges at $\mathcal{H}^{ab}_{i}$ and $\mathcal{H}^{bc}_{j}$.
This defines a co-product $\Delta$ for the Kac-Moody algebra $\mathcal{A}$ 
\begin{align}
    \Delta: \mathcal{A}(\mathcal{H}^{ac}_{k})\rightarrow \mathcal{A}(\mathcal{H}^{ab}_{i})\otimes \mathcal{A}(\mathcal{H}^{bc}_{j})
\end{align}
Once we have the factorization of a primary state $\ket{\phi_{k}}$ given by $\pi^{\dagger}$, $\Delta$ gives a factorization of the excited states on $\mathcal{H}^{ac}_{k}$ since these are created by the negative modes.  Moreover the co-product provides tensor product interpretation to our factorization map which is determined by the CFT fusion rules \cite{Gaberdiel:1993td}. 
This co-product is essentially a  generalization of the Bogoliubov transformation in Minkowski space, which expresses the operators on the a global Cauchy slice in terms of linear combinations of operators acting on subregions.
Below we will make these ideas concrete by deriving an explicit formula for the factorization of descendants using this co-product. 

\subsection{Factorization via the CFT Co-product} 
\begin{figure}[h]
\centering
\includegraphics[scale=.5]{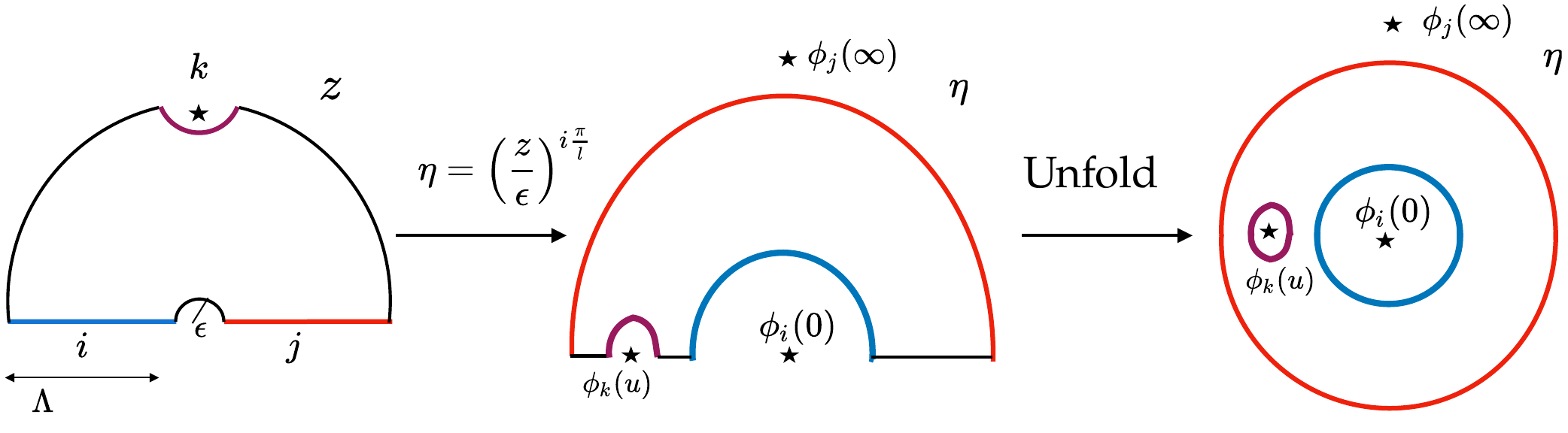}
\caption{We map the ``macaroni" geometry on the left to the 3 punctured $\eta$ plane, where we can apply the methods of \cite{Gaberdiel:1993td} to split the charge $J_{n}(u)$ using contour deformation and OPE's. Note that in the $\eta$ plane radial quantization is identified with angular quantization in the $z$ plane  }
\label{cp4}
\end{figure}

 Consider the cobordism on the left of  figure \eqref{cp4}.  
This is similar to the vacuum factorization diagram, except we have inserted a primary $\phi_{k}(z_{\infty}=\Lambda e^{i\theta })$ at the boundary surface at infinity.  We want to factorize charges $J_{n},\bar{J_{n}}$ defined around this puncture into operators acting on the two halves of the real line.  To do so we first make a conformal map to the thrice punctured upper half $\eta$ plane.  After unfolding as in the right of figure \eqref{cp4} we can derive the action of $J_{n}$ on the tensor product Hilbert space at $0,\infty$ by contour deformation. In particular, for $n>0$, we follow the method of \cite{Gaberdiel:1993td} and consider the integral around the purple contour:
\begin{align}\label{int}
\oint_{C} \frac{d\eta}{2\pi i} \,\langle \chi| (\eta- u)^{n}
J(\eta)\phi_{i}(0)\phi_{j}(\infty)|0\rangle_{k}  
\end{align} 
where $u= \eta(z_{\infty})= - \frac{\exp( i \phi) } {l}$ 
(For notational convenience we omit the boundary labels $a,b,c$ on $\phi_{i,j}$.)
Here we have taken an inner product with an arbitrary finite energy state $\ket{\chi}$ to obtain a mereomorphic function with singularities  at $0,\infty$.  The singularity structure is determined by the OPE's of $J(\eta)$ with either of the two punctures.   For example near $\eta=0$, we have the expansion
\begin{align}
    J(\eta) \phi_{i}(0) &= \sum_{m} \frac{ J_{m}(\eta=0) \phi_{i}(0) } {\eta^{m+1}}\nn
    J_{m}(0)&=\oint_{\eta=0} \eta^{m} J(\eta) d\eta
\end{align}
and near infinity where we use the coordinate $\rho=\frac{1}{\eta}$ have
\begin{align}
     J(\rho) \phi_{j}(0) &= \sum_{m} \frac{ J_{m}(\rho=0) \phi_{j} } {\eta^{m+1}}\nn
    J_{m}(\rho=0)&=\oint_{\rho=0} \rho^{m} J(\rho) d\rho \nn
\end{align}
Inserting these expansions into  \eqref{int} near each singularity and doing the integrals gives an expression of the form
\begin{align}
    \bra{\chi}J_{n} \phi_{i}(0)\phi_{j}(\infty)|0\rangle_{k}  =\sum \langle \chi|  \Delta^{i} (J_{n})   \phi_{i}(0) \Delta^{j} (J_{n}) \phi_{j}(\infty)|0\rangle_{k}
\end{align}
Since $\chi$ was arbitrary, this defines a co-product of the Kac-Moody mode $J_{n}$ which acts on the tensor product Hilbert space $\mathcal{H}_{i}^{ab}\otimes \mathcal{H}_{j}^{bc} $
\begin{align}
    \Delta(J_{n})  =\sum \Delta^{i} (J_{n})  \otimes \Delta^{j} (J_{n})
\end{align}

\begin{figure}[h]
\centering
\includegraphics[scale=.5]{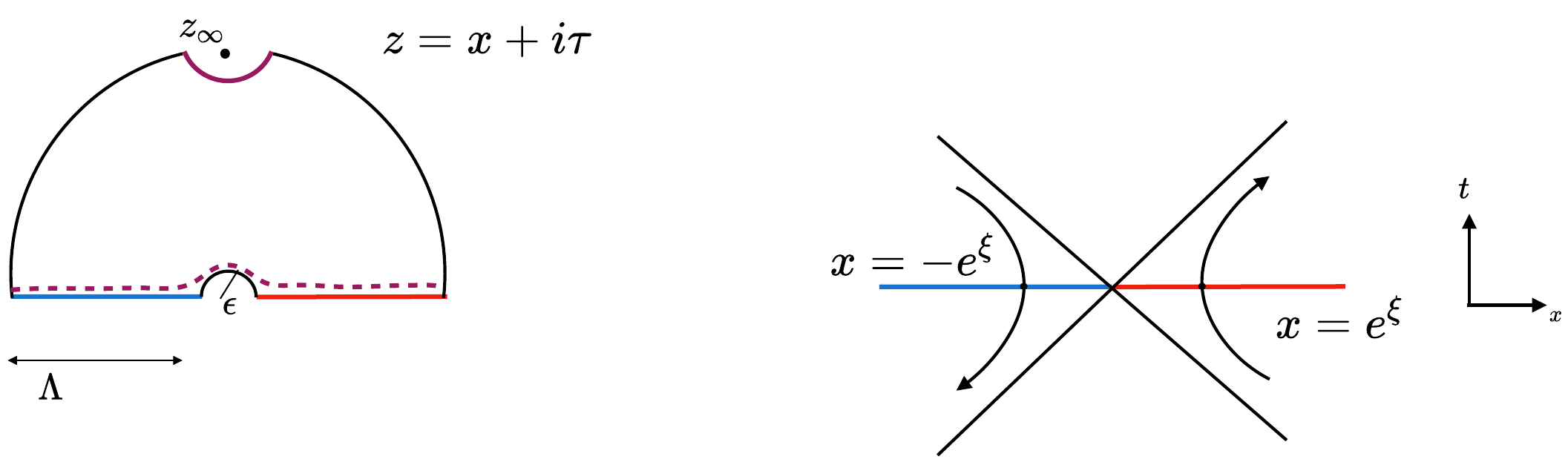}
\caption{The charge $J_{n}$ defined on the purple semicircle around $z_{\infty}$ can be deformed onto the real line.  It can then be interpreted as acting on the factorized vacuum state prepared by the ``macaroni" geometry at $\tau=0$. The co-product $\Delta(J_{n})$ obtained by contour deformation can then be related to a Bogoliubov transformation factorizing operators on a t=0  slice in Minkowski space  }
\label{Rindler}
\end{figure}

Computing this explicitly gives (for  positive $n$) 
\begin{align} \label{Delta}
    \Delta(J_{ n }(u)) = \sum_{k=0}^{n} (-u)^m\begin{pmatrix} n\\ m  \end{pmatrix} J_{m} \otimes 1  +  \sum_{q=0}^{n} (-u)^{n-q}\begin{pmatrix} n\\  q \end{pmatrix} 1 \otimes J_{-q}
\end{align}
Note that this differs slightly from the co-product formula in equation (2.9) of \cite{Gaberdiel:1993td}, particularly in the difference in the sign of the modes on the right and left region.  This is because there was a ``ket to bra" mapping hidden in the conformal transformation between the $z$ plane and the $\eta$ plane.  This is due to the logarithm in $\eta = \exp( \frac{i \pi}{l} \log z) $ , which induces a mapping $\ket{i} \rightarrow \bra{i} $ of the state on the left interval of the $z $ plane (colored blue in figure \eqref{cp4}).  
\paragraph{Comparison with Bogoliubov transformation in Minkowski space}
When mapped back to the $z$ plane in the left of figure \eqref{cp4}, the co-product formula \eqref{Delta} factorizes the annihilation operators  on a single interval Hilbert space (purple contour) into annihilation and creation operators acting on the left and right subregions of the real line.   Since this real section of the $z=x+i\tau$ plane also belongs to a $t=0$ slice of the Minkowski space (right of figure \eqref{Rindler}), we expect this factorization is related to the Bogoliubov transformation relating Minkowski and Rindler modes.  For a free Boson in Minkowski space, this is obtained by expanding the field in two ways:
\begin{align}
    \phi(x+t)|_{t=0} &= \int_{-\infty}^{\infty}  d p \, \,a_{p}e^{i p x} \quad\quad a_{-p} = a_{p}^{\dagger} \nn
    &= \int_{-\infty}^{\infty} dk \quad  \alpha_{k}^{L}\, \theta (-x)x^ {ik}  + \alpha_{k}^{R}\,\theta (x)  x^{ik}  \quad\quad \alpha_{-k}^{L,R} = \alpha_{k}^{L,R \dagger }
\end{align}
In the first line we expanded in terms of the usual Minkowski plane waves, and in the second, we expanded in Rindler plane waves  $x^{ik}=  e^{i k \xi} $ in the left and right wedges.  The corresponding creation and annihilation operators are related by 
\begin{align}\label{bog}
    p a_{p} =   \int_{0}^{\infty} d k  \frac{p^{i k}e^{\frac{\pi k }{2}}}{ \Gamma (  i k ) } \alpha_{-k}^{L} +    \int_{0}^{\infty} d k\,\, \frac{p^{-i k}e^{-\frac{\pi k }{2}}}{ \Gamma ( - i k ) }   \alpha_{k}^{R}  
\end{align}

Here we show that this can be reproduced from the co-product formula \eqref{Delta} in an appropriate limit.

To make the comparison we first take the $\epsilon \rightarrow 0 $ limit,  so that the intervals in figure (\ref{cp4}) are half infinite and can be identified with a spatial slice of the left or right Rindler Wedge.   In this limit $u\rightarrow -1$ so it drops out of the co-product formula. 

Next we take the limit  $n\gg q$, $n\gg m$, in the binomial coefficients of \eqref{cp4} in which 
\begin{align}
    \begin{pmatrix} n\\ m  \end{pmatrix} \rightarrow \frac{n^{m}}{\Gamma(q+1)} 
\end{align}

Finally we make a Wick rotation of the mode numbers:
\begin{align} \label{eq:rotate_k}
    n&\rightarrow -ip \nn
    m&\rightarrow  ik\nn
    q &\rightarrow - i k \nn
    \begin{pmatrix} n\\ m  \end{pmatrix} \rightarrow \frac{n^{m}}{\Gamma(q+1)} &\rightarrow \frac{p^{ik}e^{\frac{- \pi k}{2}}}{-ik \Gamma (-ik) }
\end{align}
so we find 
\begin{align}
    \Delta(J_{-ip}(u)) =  \int_{0}^{\infty} d k\,\, \frac{p^{i k}e^{\frac{\pi k }{2}}}{ \Gamma (  i k ) }  \frac{J_{ik}\otimes 1}{ik} \quad + \int_{0}^{\infty} d k\,\, \frac{p^{-i k}e^{\frac{-\pi k }{2}}}{ \Gamma ( - i k ) }\frac{1\otimes J_{ik}}{- ik} 
\end{align}
which is identical to the Bogoliubov transformation \eqref{bog} after identifying 
\begin{align}
    J(x)&= \pd_{x} \phi(x,t=0)\nn 
    \alpha^{L}_{-k}&= i k J_{i k } \otimes 1\nn
     \alpha^{R}_{k}&=  1 \otimes i k J_{i k } 
\end{align}

We give a heuristic argument for taking the large $n$ limit which is necessary in the comparison with known results of the Unruh effect. To begin with, one would like to deform the circle surrounding the point at infinity to a path that almost coincides with the real line, wrapping the bottom boundary of the macaroni in figure \eqref{cp4}. 
In terms of the $z$ coordinates, the mode expansion 
\begin{align}
(\eta - u)^n = (-u)^n (1- (z/\Lambda)^{i\pi/l})^n = (-u)^n \exp(n \ln(1- (z/\Lambda)^{i\pi/l}))
\end{align}
This on first sight does not look like plane waves in the Euclidean plane, which are eigen-basis with which the analysis of Unruh was based on. 
Inspecting the absolute value of the integrand, it is given by
\begin{align}
|(1- (z/\Lambda)^{i\pi/l}) | = 2 \sin(\frac{\pi \chi}{2l}), \qquad \chi = \ln \frac{z}{\Lambda}.
\end{align}
Clearly, the maximum value is located at $\chi = l$, which is equivalent to $z = \epsilon$.
Since $\sin(\pi\chi/(2l)) \leq 1$ and monotonically decreasing all the way until the IR cutoff at $z = \Lambda$, therefore  suppose we take the large $n$ limit, the norm of the integrand would fall off very rapidly. The
main contribution has to come from the region $z= \epsilon$ where $\sin(\pi\chi/(2l)) =1$.

In that region
\begin{align}
(-u)^n (1- (z/\Lambda)^{i\pi/l})^n =  (2i u)^n e^{i n \pi/(2l) z} \sin^n(\pi z/(2l))  \approx  (2i u)^n e^{i n \pi/(2l) z} + \mathcal{O}((\delta z/ \epsilon)^2).
\end{align}

We reckon the main contribution to the contour integral along the full real line (with cutoff size $2l$ in our geometry) indeed takes the form of a plane wave with momentum $n \pi/(2l)$, as desired. 

We would also like to comment on the common Wick rotation of the momenta to purely imaginary values in (\ref{eq:rotate_k}). Recall that the Unruh effect is derived in the Minkowskian signature, whereas we are working in the Euclidean signature. This suggests that this is a rather non-standard while entirely legal continuation to flip the signature -- i.e. the spatial coordinate in both the Rindler and Minkowski frames are Wick rotated. To ensure that the modes stay oscillatory, the momenta should take purely imaginary values.

\subsection{Factorization and co-product of negative modes}
We showed above that the co-product of positive Kac-Moody modes $J_{n}$ with $n>0$ analytically continues into the Bogoliubov transformation in Minkowski space, which factorizes the annihilation operators $a_{n}$.  The factorization of the creation operators $a_{-n}$ can be obtained by taking the adjoint of the RHS in the co-product formula \eqref{Delta}.  This in turn defines a factorization of the descendant states obtained by applying these creation operators to the vacuum. 

However a puzzle arises when we consider the CFT co-product formula for the negative modes, which is normally what we mean by ``the creation operators" : 
 \begin{align}
     \Delta(J_{-n}) = \sum_{m=0}^{\infty} (-1)^{m}(-u)^{-(n+m)} \begin{pmatrix} n+m-1 \\ m \end{pmatrix}  
J_{m}(\eta=0)  \otimes 1+ \sum_{l=n}^{\infty} \begin{pmatrix} l-1\\ l-n  \end{pmatrix}(u)^{l-n}1\otimes J_{l}(\eta=\infty)
 \end{align}
 This is certainly not the Euclidean adjoint of $\Delta(J_{n}) $ , so it would seem the co-product does not commute with taking the adjoint. 
 \begin{align}\label{ad}
     \Delta(J_{n}^\dagger) \neq \Delta (J_{n}) ^{\dagger}  \quad n>0 
 \end{align}
 
 The problem is that the adjoint operation in which we send $n\rightarrow  -n $ on the LHS of this equation  is not the same as the adjoint defined on the RHS by flipping the signs of the modes in each tensor factor.   To see this recall that in a Euclidean CFT, the adjoint of an operator is defined to include a time reversal operation:  in radial quantization that we are using, this time reversal is an inversion about the circle on which the Hilbert space is defined.  The adjoint thus depends on a choice of time slicing of the Euclidean plane.
 
 In particular,the time slicing defined for the Hilbert spaces associated with $J_{n}(u)$ (purple circles in figure \eqref{cp4} are not compatible with the radial time slices around $\eta=0,\infty$,  on which $J_{m} (\eta=0) \otimes 1 $ and  $1 \otimes J_{l} (\eta= \infty )  $  are defined.   We should thus label our adjoint operations with a choice of time slice, in which case equation \eqref{ad} becomes less mysterious.  Notice that Euclidean adjoint for the charges around $\eta=0,\infty $ involve radial inversion in the $\eta$ coordinate system which corresponds to inversion in the angular coordinate on the $z$ plane.  
 When analytically continued to  Minkowski space, angular inversion becomes  reversal of the Rindler time, which is compatiable with the global reversal of Minkowski time.  This is why the Euclidean adjoint applied to the RHS of the co-product \eqref{Delta} is compatible with the adjoint in Minkowski space. 
\subsection{A local tensor product structure for CFT}
 Despite it's unusual conformal frame, the factorization cobordism on the left of figure \eqref{cp4} is conformally equivalent to a three point function of boundary primary operators, as shown by the mapping to the $\eta$ plane.  This is a linear operator  $\pi$ that defines fusion in a CFT as in \eqref{pi}, which can be interpreted as a way of decomposing tensor product representations of the chiral algebra into irreducibles.  More precisely, $\pi$ is an intertwiner with respect to the co-product $\Delta$:
 \begin{align}
     J_{n} \pi& = \pi \Delta (J_{n })
 \end{align}
 However the relevant tensor product on which $\Delta(J_{n})$ is well defined is not the usual tensor product of Hilbert spaces, which would not preserve the central charges of the Kac-Moody Algebra\footnote{The central charges  of the chiral algebra would just add under the normal tensor product of vector spaces}. Instead a quotient must be imposed to define a tensor product that preserves central charges and is compatible with the fusion rules of a CFT \cite{Gaberdiel:1993td} \cite{Moore:1988qv}. 
 
Let's consider this quotient from the point of view of Hilbert space factorization.  Naively,  the Hilbert space of two intervals has twice as many states as that of one interval, so identifying these Hilbert spaces requires a quotient of some sort. For example, in the zero mode sector of the compact boson this was implemented by our gluing rules and the sum over edge modes, which projects onto an entangled subspace.  However there remains two independent towers of states that we can create on top of this subspace via the Kac-Moody modes on either interval.  This is related to the fact that the central charge would double should we take the usual tensor product on the two intervals.  However the aforementioned quotient which defines the fusion rules cuts down the states so that one can fuse two intervals into one.  This suggests that the combination of the quotient on the zero modes and the excited states give a CFT analog of the entangling product that was defined for gauge theories \cite{Donnelly:2016auv}.

\begin{figure}[h]\label{quotient}
\centering
\includegraphics[scale=.5]{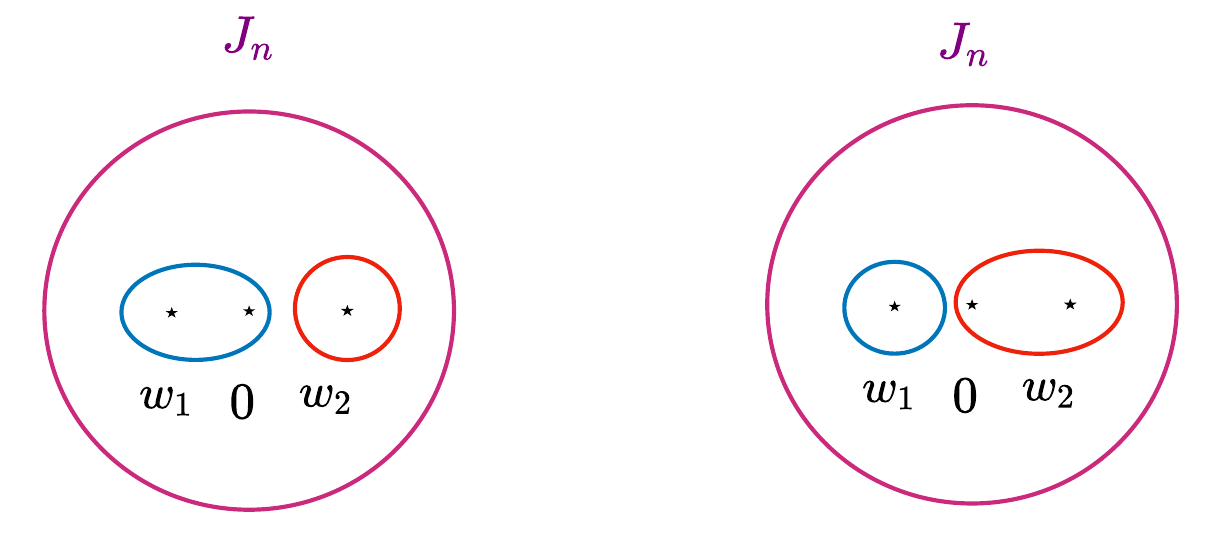}
\caption{Two ways of computing the co-product for negative modes, corresponding to two choices of contour deformations and OPE's}
\end{figure}

For completeness, let us briefly recall the definition of the CFT quotient \cite{Gaberdiel:1993td}. Consider a mapping $w(\eta)$ of the thrice punctured $\eta$ plane, taking the point $\eta= u,0,\infty$ to $w=\infty,w_{1}, w_{2}$. To determine the co-product $\Delta$, we derive the action of $J_{n}$ on the tensor product Hilbert space at $w_{1},w_{2}$ by computing an integral similar to \eqref{int} but for negative modes: 
\begin{align}
\oint_{C} \frac{dw}{2\pi i} \,\langle \chi| w^{-n}
J(w)\phi_{i}(w_{1} ) \phi_{j}(w_{2})|0\rangle  \quad \quad n>0\nn
\end{align} 
Where $C$ is  a large contour around infinity (purple circle in figure \eqref{quotient} ).   There is now a pole at $w=0$ whose residue can be evaluated either by taking the OPE of $J(w)$ with primaries inserted at $w_{1}$ or at $w_{2}$. 
The two choices of OPE's give two different expressions for the co-product for negative modes:
\begin{align}
    \Delta^{1}(J_{-n}) &=  \sum_{l=n}^{\infty} \begin{pmatrix} l-1\\ n-1\end{pmatrix} (-w_{1})^{l-n} J_{-l}\otimes  1+\sum_{m=0}^{n} \begin{pmatrix} n+m-1\\ m\end{pmatrix} (-1)^{m}w_{2}^{-(n+m)}1 \otimes  J_{m}  \nn
    \Delta^{2}(J_{-n}) &= \sum_{m=0}^{n} \begin{pmatrix} n+m-1\\ m\end{pmatrix} (-1)^{m}w_{1}^{-(n+m)} J_{m} \otimes 1+ \sum_{l=n}^{\infty} \begin{pmatrix} l-1\\ n-1\end{pmatrix} (-w_{2})^{l-n}1\otimes J_{-l}
\end{align}
The quotient is defined by the equivalence relation on the tensor product space that sets  $\Delta^{1}(J_{-n})=\Delta^{2}(J_{-n}) $.  This defines the fusion product
\begin{align}
    \mathcal{H}^{ab}_{i}\boxtimes  \mathcal{H}^{bc}_{j}=\mathcal{H}^{ab}_{i}\otimes  \mathcal{H}^{bc}_{j}\big/\{ \Delta^{1} = \Delta^{2}\}
\end{align}
on which the co-product $\Delta$ is well defined. 
 \section{Conclusion}
  
In this work we proposed an extension of a 2D CFT which gives a factorization of the Hilbert space in terms of OPE data.   To identify the entanglement boundary state in this frame work, we introduced a constraint called the E brane axiom and proposed the vacuum Ishibashi state as a solution.   This leads to CFT edge modes corresponding to super-selection sectors labelled by physical boundary conditione, i.e. D branes.   In each superselection sector, we related the factorization to the co-product formula of the chiral symmetry algebra. This co-product defines a tensor product structure compatible with fusion, and we conjecture that when combined with the sum over superselection sectors this gives the analogue of the entangling product for CFT's. 

This paper is a preliminary step in understanding the nature of CFT edge modes, and we have focused our analysis on free bosons in Euclidean signature.  It would be interesting to generalize to other radii of the free boson \footnote{ At rational and self dual radii,  the spectrum and the space of conformal boundary conditions enlarges beyond the usual Dirichlet and Neumann cardy states.  It seems all the boundary conditions have been classified in  \cite{Gaberdiel:2001zq}.   
} where new boundary conditions emerge, and to more general theories such as RCFTs with non-abelian fusion algebras and holographic CFT's.  In the latter case, the entropy as given by the Ryu-Takayanagi formula has a leading area term which has been conjectured to arise from a sum over gravitational edge modes \cite{Lin:2017uzr}.  However the nature of the edge modes remain mysterious both in the bulk gravity theory and in the boundary CFT.  Perhaps our proposal for the CFT factorization, combined with elements of AdS/BCFT \cite{Fujita:2011fp}, will shed light on this problem.

It should be possible to connect our Euclidean boundary state approach to a direct derivation of the CFT edge modes in Minkowski signature.   For example, in the case of the compact boson, the edge mode on a circle was derived from a direct analysis of the symplectic potential \cite{Freidel:2017nhg,Freidel:2017wst}.  In these references, the closed string worldsheet is treated as a strip with time like boundaries corresponding to the branch cut for the winding modes.  As a result, one finds that the symplectic potential contains  ``corner term" which indicates that $\tilde{\phi}$ is an edge mode degree of freedom for $\phi$.
In our corbodism picture, the same kind of edge modes arise from the ``zipper" cobordism in \eqref{fcob} which maps states on the circle to states on an interval.  As in \cite{Freidel:2017nhg,Freidel:2017wst}, we find that  $\tilde{\phi}$ as an edge mode in the Neumann extension.

\begin{figure}[h]
\centering
\includegraphics[scale=.45]{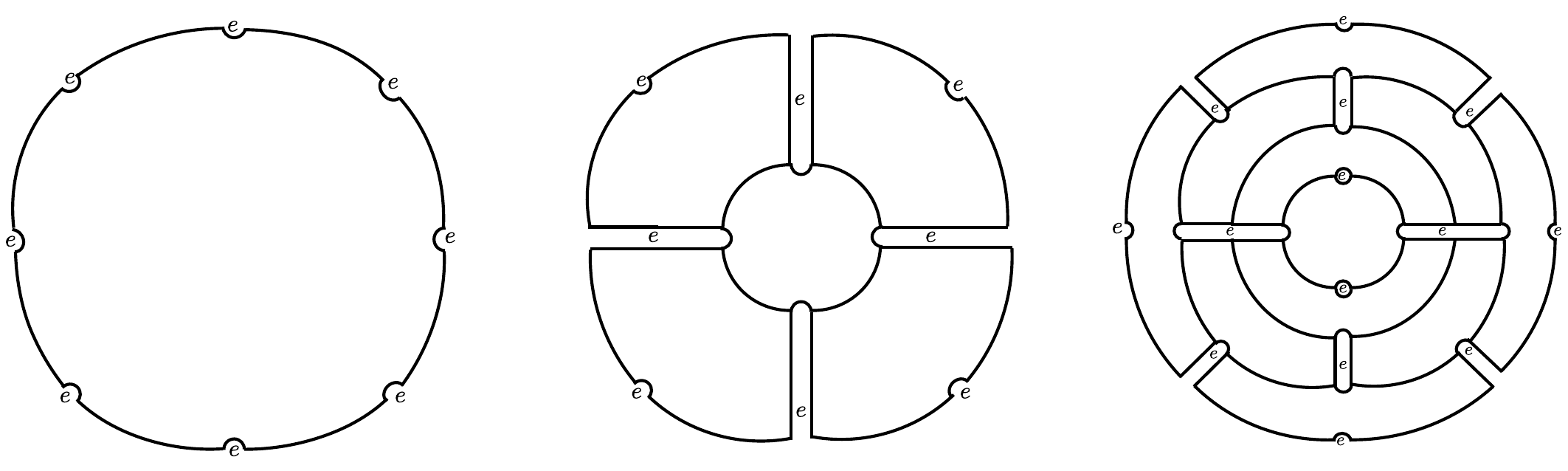}
\caption{Factorization of the circle into many intervals is given by the corbodism on the left.  The same factorization map can be achieved by the a network of local cobordisms as in the middle and right figure.  The E brane boundary condition implies these networks give an arbitrarily good approximation to the original state}
\label{Disk}
\end{figure} 

Finally we would like to point out a connection between our work to tensor network renormalization \cite{Milsted:2018vop} and the recently proposed picture of a CFT as being composed of entangled ``BC bits"\cite{VanRaamsdonk:2018zws}.  These connections can be understood by considering how to factorize a CFT state on the circle into multiple intervals.  The factorization cobordism involves a disk with  many semi circle regularizations with E brane conditions at the boundary as in the left of figure \eqref{Disk}.  This cuts the circle into many intervals, and the resulting path integral gives the factorization map.  However, when we split the circle into many intervals, the modular evolution is complicated by presence of singularities that cut and reglue the intervals\cite{Wong:2018svs} \cite{Donnelly:2018ppr}.  Thus we no longer have a simple local operator like $L_{0}$ as the modular Hamiltonian.  However there is another way to perform the factorization which makes use of locality and symmetry of the problem.

\begin{figure}[h]
\centering
\includegraphics[scale=.45]{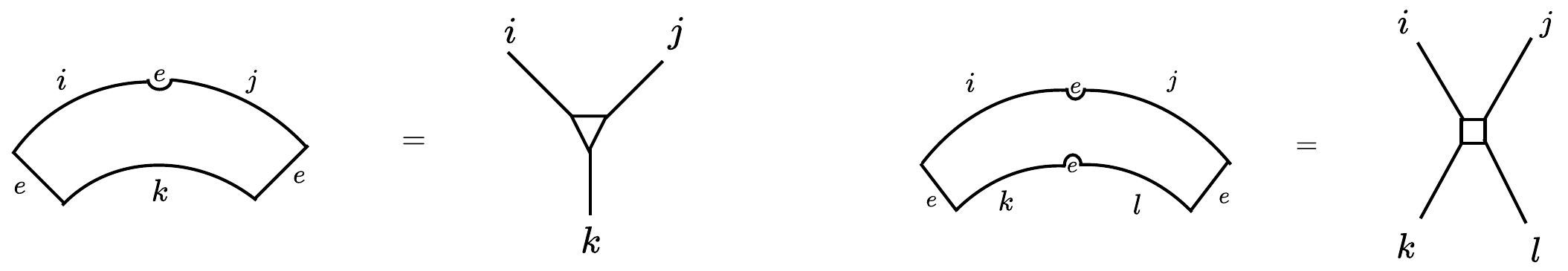}
\caption{Local cobordisms used to build up the CFT state on a circle.  These are continuum analogs of the coarse graining isometry and the disentangler of the MERA tensor network.  Indeed each cobordism is an infinite dimensional tensors whose components deterimned by the OPE coefficients of the CFT.}
\label{Ltensors}
\end{figure}

Instead of preparing and factorizing the state in one go, we do so layer by layer using local cobordisms that split one interval into two as in the left diagram of figure \eqref{Ltensors} . Thus we can build up the state via a network composed of a ``coarse graining'' cobordism\footnote{ in \cite{Shiozaki:2016cim} a similar TQFT cobordism network was built for symmetry protected topological (SPT) phases in 1 spatial dimension.  Such a network was shown to be equivalent to the MPS description of SPT phases.}  (middle of figure \eqref{Disk}).
Such a network will give an arbitrarily good approximation to the original path integral, because the geometry differs only in small holes and narrow slits where we have imposed the E brane condition.  This gives a precise formulation of the proposal in \cite{VanRaamsdonk:2018zws} to describe a CFT in terms of entangled ``BC bits", which corresponds to our coarse graining cobordism.

The coarse graining procedure described above allows entanglement to accumulate, as can be seen by the long slits that go deep into the center of the disk.   As an alternative we can introduce the ``disentangling" cobordisms, which are two to two scatterings of open strings.  As shown in the right of figure \eqref{Disk}, this reproduces the original state, up to small holes and slits that now extend only over a single layer.
The resulting network is structually very similar to the MERA tensor network.  Indeed recent numerical work has shown striking similarities between the MERA network and the Euclidean path integral \cite{Milsted:2018san} \cite{Milsted:2018yur} .  Our local cobordism description of the CFT provides a possible rationale for this similarity.   In particular, there is a natural truncation of the local cobordisms in figure \eqref{Ltensors} into finite dimensional tensors.  Since each local cobordism implements a local Dilatation, we expect that the descendants will be subleading relative to to the primaries.  Thus to first order, we can express each cobordism in terms of the OPE coefficients of the primary fields.  It would be interesting to see if this truncation gives a useful simulation of the CFT state.  

\section{Acknowledgements}
GW would like to thank Matthias Gaberdiel, Ronak Soni, William Donnelly, and Laurent Freidel, Matt Headrick, for valuable discussions related to this work.  GW and LYH acknowledges the support of Fudan University and the Thousands Young Talents Program and thank Perimeter Institute for hospitality as a part of the EmmyNoether Fellowship programme. This work is supported by the NSFC grant number 11875111 and 11922502.
GW thanks Stanford Institute for Theoretical Physics, Hong Kong University of Science and Technnology IAS, and the QIST conference 2019 at the Yukawa Institute for hospitality while this work was being finished. Research at Perimeter Institute is supported by the Government of Canada through the Department of Innovation, Science and Economic Development Canada and by the Province of Ontario through the Ministry of Research, Innovation and Science.

\bibliographystyle{utphys}
\bibliography{open-closed}

\end{document}